\documentclass[]{revtex4-2}

\usepackage{graphicx} 
\usepackage{amsmath}
\usepackage{bm}
\usepackage{appendix}
\usepackage{xcolor}
\usepackage{hyperref}
\usepackage{natbib}
\usepackage{orcidlink}
\usepackage{amssymb}
\usepackage{soul}

\begin{document}

\title{Multi-Dataset Inverse Problem Solving with Distributed Generative AI}

\author{Daniel Lersch\orcidlink{https://orcid.org/0000-0002-0356-0754}}
\email{dlersch@jlab.org}
\affiliation{Data Science Department, Thomas Jefferson National Accelerator Facility, Newport News, VA, USA}

\author{Steven Goldenberg\orcidlink{https://orcid.org/0000-0002-5264-6298}}
\email{sgolden@jlab.org}
\affiliation{Data Science Department, Thomas Jefferson National Accelerator Facility, Newport News, VA, USA}

\author{Johann Rudi\orcidlink{https://orcid.org/0000-0002-6563-9265}}
\affiliation{Department of Mathematics, Virginia Tech, Blacksburg, VA, USA}

\author{Markus Diefenthaler\orcidlink{https://orcid.org/0000-0002-4717-4484}}
\affiliation{Electron Ion Collider Group, Thomas Jefferson National Accelerator Facility, Newport News, VA, USA}

\author{Kevin Braga\orcidlink{https://orcid.org/0009-0008-8752-8292}}
\affiliation{Department of Physics, William $\&$ Mary, Williamsburg, VA, USA}

\author{Xingfu Wu\orcidlink{https://orcid.org/0000-0001-8150-5171}}
\affiliation{Mathematics and Computer Science Division, Argonne National Laboratory, Lemont, IL, USA}

\author{Yaohang Li\orcidlink{https://orcid.org/0000-0003-0178-1876}}
\affiliation{Department of Computer Science, Old Dominion University, Norfolk, VA, USA}

\author{Nobuo Sato\orcidlink{0000-0002-1535-6208}}
\affiliation{Theory Division, Thomas Jefferson National Accelerator Facility, Newport News, VA 23606, USA}

\begin{abstract}
    Extracting a shared set of unknown, not directly measurable quantities from multiple, heterogeneous datasets is a common challenge across scientific domains. A prominent example is the combination of datasets obtained from different measurements with different settings (e.g. varying detector resolutions or non-uniform coverage of the available feature space). Analyzing such datasets jointly, rather than independently or after naive merging, is essential for obtaining precise and unbiased estimates of the unknowns, but requires careful treatment of dataset heterogeneity and is computationally demanding. We present a generalized framework for simultaneously analyzing multiple heterogeneous datasets in the context of generative AI-based inverse problem solvers. Building on our recent Scalable Asynchronous Generative Inverse Problem Solver (SAGIPS) framework, we extend the well-established distributed data-parallel training paradigm to non-identically distributed datasets, where each dataset is controlled by the same set of unknown inference parameters but covers a different region of the available feature space. Each dataset is processed through its own forward operator and discriminator, providing complementary constraints that collectively guide a shared generator toward global parameter consistency. We validate the approach using a controlled setup inspired by a multi-detector Rutherford scattering experiment, demonstrating that the method reliably recovers the underlying physical parameters and that the resolution of inferred parameters improves as additional datasets are incorporated. We provide numerical evidence that our framework is robust to different data fidelities, which arise from unknown detector systematics in the Rutherford experiment, and we show the scaling behavior on multi-GPU leadership computing systems. The results show that our approach is well suited for real-world multi-dataset analyses in which experimental conditions vary across measurements.
\end{abstract}

\maketitle

\section{Introduction}
\label{sec:introduction}
Inverse problem solving remains a central challenge across a variety of scientific domains. This includes, but is not limited to, inferring high-dimensional quantities~\cite{Patel2022,Ray2022} or combining multiple heterogeneous, high-volume datasets to extract a set of common unknowns. The latter has been addressed in a variety of uses cases, such as geophysics~\cite{Mustafa2021, bosch2006} or inverse scattering problems~\cite{liu2025, liu2026}. Strategies proposed for these use-cases range from probabilistic coupling~\cite{bosch2006} to deep-learning-based methods~\cite{Mustafa2021}, including multi-task learning~\cite{liu2025} or multi-level architectures~\cite{liu2026}. In many of these scenarios, the measurements are performed under different conditions therefore producing datasets that each probe distinct and sometimes non‑overlapping regions of the feature space. 

Another motivating example comes from the QuantOm project\footnote{\url{https://www.anl.gov/phy/quantom}}, which aims to image the internal structure of the proton and other nuclear systems in terms of their constituent quarks and gluons. This structure is encoded in quantum correlation functions that cannot be measured directly, but must instead be inferred from high-energy electron-scattering data. This constitutes an inverse problem. Solving this inverse problem requires, in practice, combining measurements from different experiments, targets, and kinematic regions; this, in turn, provides more complete kinematic coverage, and helps disentangle contributions from different quark flavors. Given the resulting dataset sizes and computational complexity, QuantOm necessitates high-performance computing (HPC) resources, leading to two fundamental challenges: efficiently parallelizing the scientific workflow across compute units, and combining heterogeneous datasets in a manner that preserves consistency of the shared physical parameters.

We address these challenges by extending the SAGIPS~\cite{SAGIPS_2025} (Scalable Asynchronous Generative Inverse Problem Solver) framework to enable simultaneous, multi-dataset inverse analysis using distributed generative modeling. The key idea is to use Generative Adversarial Networks (GANs)~\cite{goodfellow2014generative,salimans2016improved} as a flexible hypothesis generator, while distributing the dataset-specific forward operators and discriminators across multiple GPUs. This allows each dataset to contribute independent constraints that are incorporated through shared generator gradient updates. In spirit, our approach shares the underlying philosophy of multi-task learning strategies such as~\cite{liu2025}, treating each dataset as a distinct source of constraints toward a shared reconstruction. Unlike these approaches, we synchronize a single shared generative model across all datasets using distributed data parallelism. This design choice follows naturally from SAGIPS's existing architecture~\cite{SAGIPS_2025}, without requiring fundamental changes to the underlying distributed-training paradigm.

To demonstrate and validate our method, we construct a hypothetical experiment inspired by the QuantOm project and the challenges it poses for multi‑dataset inverse analysis. Specifically, we design a multi-detector Rutherford scattering setup, where the underlying physical laws are well known. This provides a clear benchmark for assessing whether the proposed framework can recover the true physical parameters and how performance improves as more datasets are incorporated. The example exhibits the challenges of real scientific scenarios where each experimental configuration probes a different part of the physics process and may suffer from (unknown) systematic effects. We briefly examine the feasibility of extending the multi-dataset analysis to a more complex problem, where the generator is tasked with recovering two unknown density distributions from three simultaneous datasets. As this is an exploratory feasibility study rather than a central component of this work, the results are presented separately in appendix~\ref{sec:app_proxy2d}.

The contributions of this work are as follows:
\begin{enumerate}
\item We extend the distributed data-parallel paradigm to non-identically distributed datasets by demonstrating that gradient aggregation remains valid when all datasets share a common set of unknown parameters.
\item We propose a scalable multi‑dataset workflow that associates each dataset with its own forward operator and discriminator, while synchronizing generator gradients across all GPUs.
\item We show the robustness with respect to multiple gradient-transport mechanisms to confirm that our findings are not specific to a single gradient communication strategy. 
\item We demonstrate the performance and scalability of the multi-dataset approach on single-node and multi-node HPC systems with up to 120 GPUs. 
\end{enumerate}
Our results show that the proposed framework successfully recovers the unknown physical parameters, improves resolution as datasets are added, and maintains robustness in the presence of unknown detector systematics. Beyond the synthetic Rutherford benchmark, this approach is broadly applicable to multi‑dataset inverse problems in physics and other scientific domains.

\section{Problem Formulation and Proposed Approach}
\label{sec:theory}
\begin{figure}[htbp]
 \centering
 \includegraphics[width=0.6\textwidth]{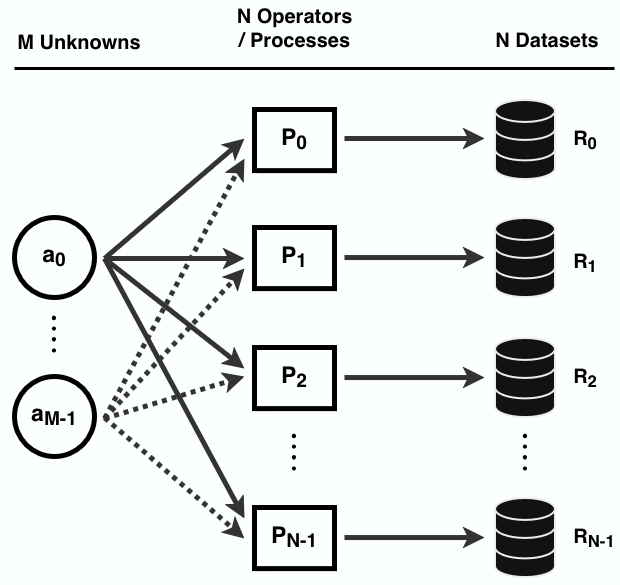}
 \caption{A large volume dataset consists of $N$ subsets $R_{0}$,...$R_{N-1}$. Each of the subsets is connected via an operator $P_i$ to a set of M unknowns: $R_i = P_i[a_{0},\ldots,a_{M-1}]$.}
 \label{fig:fig_1}
\end{figure}

In this work, we consider the inverse problem presented in Fig.~\ref{fig:fig_1}, in which $N$ independent measurements yield $N$ datasets $R_0, \ldots, R_{N-1}$, each probing a different region of the feature space. 
All datasets depend on a shared set of $M$ unknowns $a_0, \ldots, a_{M-1}$ that govern the underlying physical process and can be represented as numerical values or multi-dimensional tenors. These unknowns cannot be measured directly, but the forward process (or operator) $P_i$ associated with each dataset $R_i$ is assumed to be known and provides a mapping from the unknowns to the observations:
\begin{equation}
\label{def_R_i}
R_i = P_i[a_0, \ldots, a_{M-1}], \quad i = 0,\ldots, N-1.
\end{equation}
Each operator may incorporate experiment-specific conditions and is, in general, not analytically invertible. The crux of this problem is that the datasets may probe the available feature space differently, either due to differences in measurement configuration, limited sensitivity of the experimental equipment, or differences in the underlying physics processes or kinematic regimes being covered. As a result, each dataset carries distinct information, making the collection complementary rather than redundant. Moreover, the combined size of the datasets and the computational cost of the forward operators often exceed the resources of a single GPU, necessitating a distributed approach. To address these constraints, we build on the SAGIPS generative inverse solver~\cite{SAGIPS_2025}, which is designed to recover the unknown $x$ from the forward process $P(x)=y$, given the observed data $y$. SAGIPS' core optimization paradigm can be formally expressed as: 
\begin{equation}
    \label{def_sagips}
    \phi^{\ast} = \mathrm{arg}\min_{\phi}\Big\{F_{\psi}[P(x_{\phi}),y] \Big\} \implies x^\ast = x_{\phi^{\ast}} 
\end{equation}
where $x_\phi$ represents the unknowns predicted by a model with learnable parameters $\phi$. $P$ is the computational realization of the forward process that translates $x_\phi$ such that it can be directly compared to the measured data $y$. The objective $F_{\psi}$ is agnostic to any ordering in $y$ or $P(x_{\phi})$ and may itself contain trainable components $\psi$. The optimum solution to Eq.~\ref{def_sagips} is denoted by $x^{\ast}$. It should be noted that, depending on the underlying objective $F$, Eq.~\ref{def_sagips} may also be expressed as a maximization. For the problem discussed in this work, we denote the predicted unknowns as $x_\phi = (A_0, \ldots, A_{M-1})$, using capital letters, versus their true counterparts $a_j$ in Fig.~\ref{fig:fig_1}. While SAGIPS supports different means to solve Eq.~\ref{def_sagips}, the current implementation uses a GAN~\cite{goodfellow2014generative}, therefore we need to solve the coupled optimization problem:
\begin{align}
    \label{def_sagips_gan1}
    \psi^\ast &= \mathrm{arg}\min_{\psi}\Big\{ \mathcal{L}[D_{\psi}(P(G_{\phi}(n)),0] + \mathcal{L}[D_{\psi}(y), 1] \Big\}
    & \implies D^\ast &= D_{\psi^\ast}, \\
    \label{def_sagips_gan2}
    \phi^\ast &= \mathrm{arg}\min_{\phi}\Big\{ \mathcal{L}[D_{\psi}(P(G_{\phi}(n)),1] \Big\}
    & \implies G^\ast &= G_{\phi^\ast},
\end{align}
where $D^*$ and $G^*$ denote the optimal discriminator and generator networks respectively, with $D_{\psi}$ and $G_{\phi}$ representing their parametric forms during optimization. The input tensor $n \sim \mathcal{N}(0, I)$ represents a noise vector sampled from a standard normal distribution. The loss function $\mathcal{L}$ is usually set to be the binary cross entropy~\cite{goodfellow2014generative}, although mean squared error is also commonly employed~\cite{MaoLiXieEtAl17}. It should be noted that $(D^*, G^*)$ represents the Nash equilibrium of the coupled (iterative) optimization in Eqs.~\ref{def_sagips_gan1} and~\ref{def_sagips_gan2}. The generator can only learn expressive patterns if the discriminator is sufficiently well-trained, and vice versa.

\subsection{Distributed Multi-Dataset Analysis}
\label{sec:theory_dmda}
The backbone of the presented method is the distributed data parallel training formalism which is discussed in Appendix~\ref{sec:app_ddp}. 
\begin{figure}[htbp]
    \centering
    \includegraphics[width=1.0\textwidth]{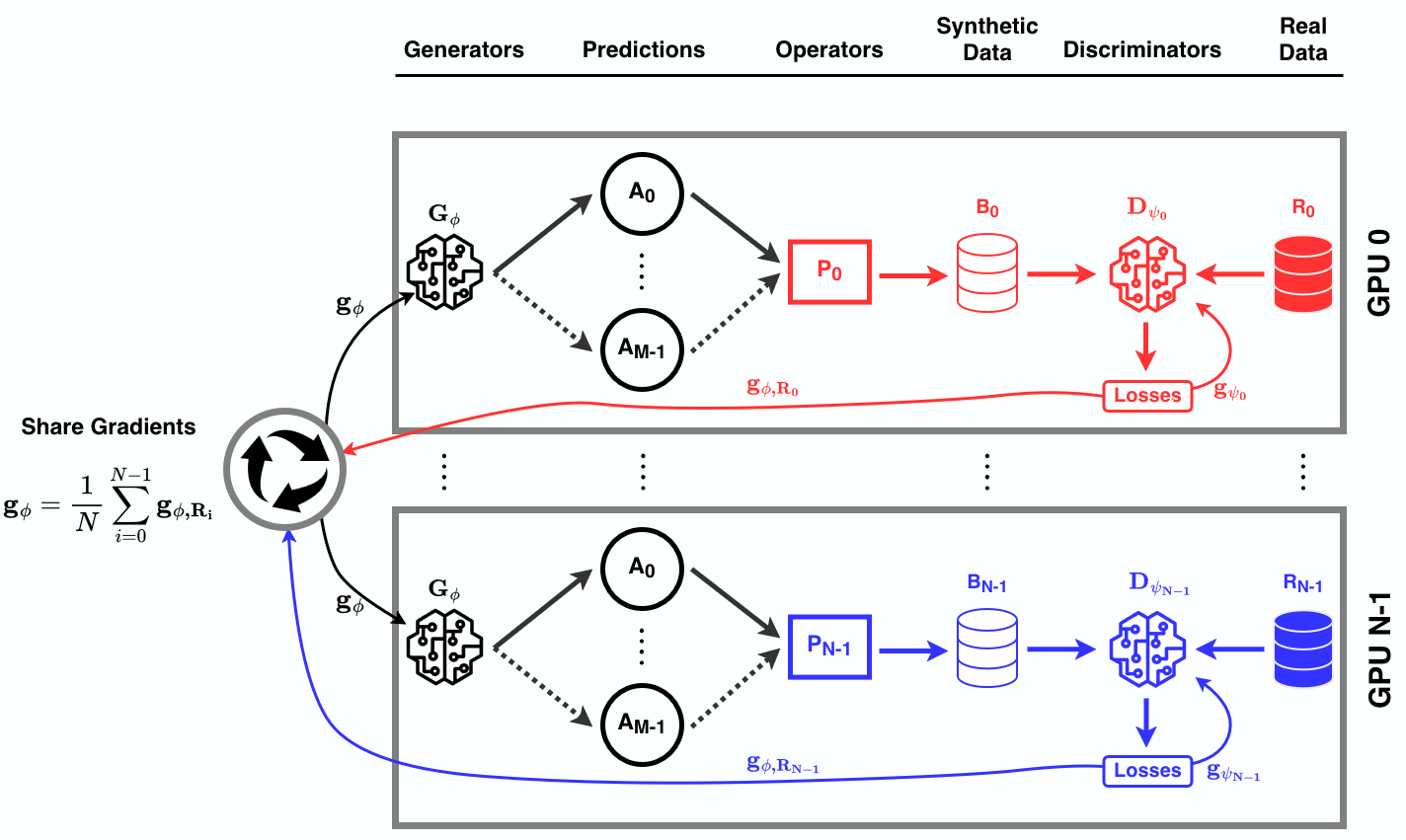}
    \caption{Schematic presentation of the proposed approach to analyze $N$ datasets in parallel through the SAGIPS inverse problem solver. Displayed here are, for simplicity, the first and last out of $N$ GPUs. The red (blue) colors indicate the operator, discriminator and real dataset that are bound to the first (last) GPU. The generator is identical on each GPU and updated through the shared gradients $g_\Phi$. The discriminator on each GPU is trained locally and thus shares on information with adjacent GPUs.}
    \label{fig:fig_2}
\end{figure}
The key idea is to assign each dataset $R_i$ to a dedicated GPU $i$ along with its own forward operator $P_i$ and discriminator $D_{\psi_i}$. A visual presentation of this is displayed in Fig.~\ref{fig:fig_2}, where a noise vector $n_i \sim \mathcal{N}(0, I)$ is passed to the local replica of the generator $G_{\phi}$ to produce predictions:
\begin{equation}
 \label{def_G_pred}
 G_{\phi}(n_i) = (A_0,\ldots,A_{M-1}),
\end{equation}
which are passed through the dataset‑specific operator $P_i$ to yield synthetic data:
\begin{equation}
    \label{def_B_i}
    B_i = P_i[A_0,\ldots,A_{M-1}].
\end{equation}
The synthetic and real data are then fed to the local discriminator $D_{\psi_i}$. The optimization problem in~\ref{def_sagips_gan1} and~\ref{def_sagips_gan2} is solved with alternating stochastic gradient descent steps for the discriminators $D_{\psi_i}$ and generator $G_\phi$. The gradients are derived using the calculation in Eq.~\ref{def_grad} from Appendix~\ref{sec:app_ddp}. Each GPU locally determines the discriminator gradients $g_{\psi_i}$:
\begin{equation}
    \label{def_grad_D}
    g_{\psi_i} = \nabla_{\psi_i}\Big\{\mathcal{L}[D_{\psi_i}(B_i),0] + \mathcal{L}[D_{\psi_i}(R_i), 1] \Big\}
\end{equation}
and the generator gradients $g_{\phi,R_i}$ respectively:
\begin{equation}
    \label{def_grad_G}
    g_{\phi,R_i} = \nabla_{\phi}\Big\{\mathcal{L}[D_{\psi_i}(B_i),1] \Big\}
\end{equation}
Eg.~\ref{def_grad_G} does not directly show a dependency to $R_i$, but to the discriminator $D_{\psi_i}$ that has been trained on the dataset $R_i$, following Eq.~\ref{def_grad_D}. The discriminator gradients $g_{\psi_i}$ remain local to each GPU and are used to update the dataset-specific discriminator parameters $\psi_i$ independently. The generator gradients $g_{\phi,R_i}$ however, are shared and aggregated via summation across all GPUs into synchronized gradients $g_{\phi}$ as described in Appendix~\ref{sec:app_ddp}. This ensures that all replicas of $G_{\phi}
$ produce predictions that are simultaneously consistent with all datasets $R_0,\ldots,R_{N-1}$.

\subsection{Validity of Non-IID Gradient Accumulation}
\label{sec:theory_validity}
In standard distributed data-parallel (DDP) training, gradient averaging across GPUs requires that local data shards be identically distributed (IID) so that each gradient represents an unbiased estimate of the same expectation. In the proposed framework, the IID requirement is not imposed on the datasets $R_0,\ldots,R_{N-1}$, because gradient accumulation remains valid regardless (see Appendix~\ref{sec:app_ddp_multi}, Eq.~\ref{def_G_joint}). However, for this aggregation to be scientifically meaningful, all datasets must be governed by the same underlying unknowns $a_0,\ldots,a_{M-1}$, i.e. the generator parameters $\phi$ must correspond to one consistent set of physical quantities across all datasets. In the particular case of a GAN, the generator gradients do not depend explicitly on the dataset $R_i$ (see Eq.~\ref{def_grad_G}), but implicitly depend on $R_i$ through the parameters $\psi_i$ of the locally trained discriminator $D_{\psi_i}$, which encodes information about $R_i$ through its training process. This statement holds as long as the discriminator has sufficient capacity to represent the distinguishing features of $R_i$ and is trained sufficiently long to encode the relevant dataset structure.
Each discriminator therefore contributes gradients that encode its dataset's constraints on the shared unknowns, and aggregating these gradients yields updates that reflect the joint information of all datasets simultaneously. Conceptually, this resembles multi-task learning: even when tasks are heterogeneous, they can jointly train a shared representation as long as they depend on common latent variables. A potential non-IID nature of the datasets therefore enhances, rather than invalidates, the learning process.

\subsection{Gradient Alignment}
\label{sec:theory_align}
We considered gradient alignment which is commonly employed in multi-task learning to detect and mitigate conflicting gradient directions~\cite{yu2020gradient}. However, three fundamental issues make such methods unsuitable for the proposed framework. First, the opening angle between gradients is not a reliable indicator of harmful updates — gradient disagreement may simply reflect complementary experimental constraints rather than detrimental interference, as argued in Section~\ref{sec:theory_validity}. Second, monitoring pairwise gradient angles across all GPUs introduces a computational overhead that scales poorly with the number of datasets, negating the efficiency gains of distributed training. Third, incorporating gradient alignment in a manner consistent with the DDP all-reduce mechanism — without introducing synchronization bottlenecks or asynchronous updates — is non-trivial and would require fundamental modifications to the training infrastructure. For these reasons, gradient alignment is not pursued in this work and remains an open problem for future investigation.

\subsection{Further Data Sharding and Multiplicity}
\label{sec:theory_multiplicity}
The approach presented in Fig.~\ref{fig:fig_2} extends naturally to distributing each dataset $R_i$ across $m$ GPUs, requiring $N \times m
$ GPUs in total, with standard DDP applied within each dataset group independently. This allows reducing the computational load per GPU, as it only sees a fraction of the initial size of $R_i$. From this point on, we will refer to $m$ as the multiplicity. For example, a multiplicity $m=2$ represents a scenario where each of the $N$ datasets is divided across two GPUs. 

\subsection{Potential Failure Modes}
\label{sec:theory_failure}
It should be noted that the validity of the proposed framework relies on the assumption that all datasets $R_0,\ldots,R_{N-1}$ are governed by the same set of unknowns $a_0,\ldots,a_{M-1}$. If this condition is violated, the accumulated gradients may become contradictory and the shared generator may fail to converge to a meaningful solution. Furthermore, each dataset should be represented with sufficient statistical significance, as a statistically underrepresented dataset may result in a poorly trained local discriminator, which in turn provides uninformative or misleading gradients to the shared generator and thus potentially degrading the overall prediction quality. Detecting such failure modes in practice may require additional diagnostics, such as monitoring discriminator loss plateaus or cross‑dataset consistency checks.

\section{Experiments}
\label{sec:exp}
\begin{figure}[htbp]
    \centering
    \includegraphics[width=0.85\linewidth]{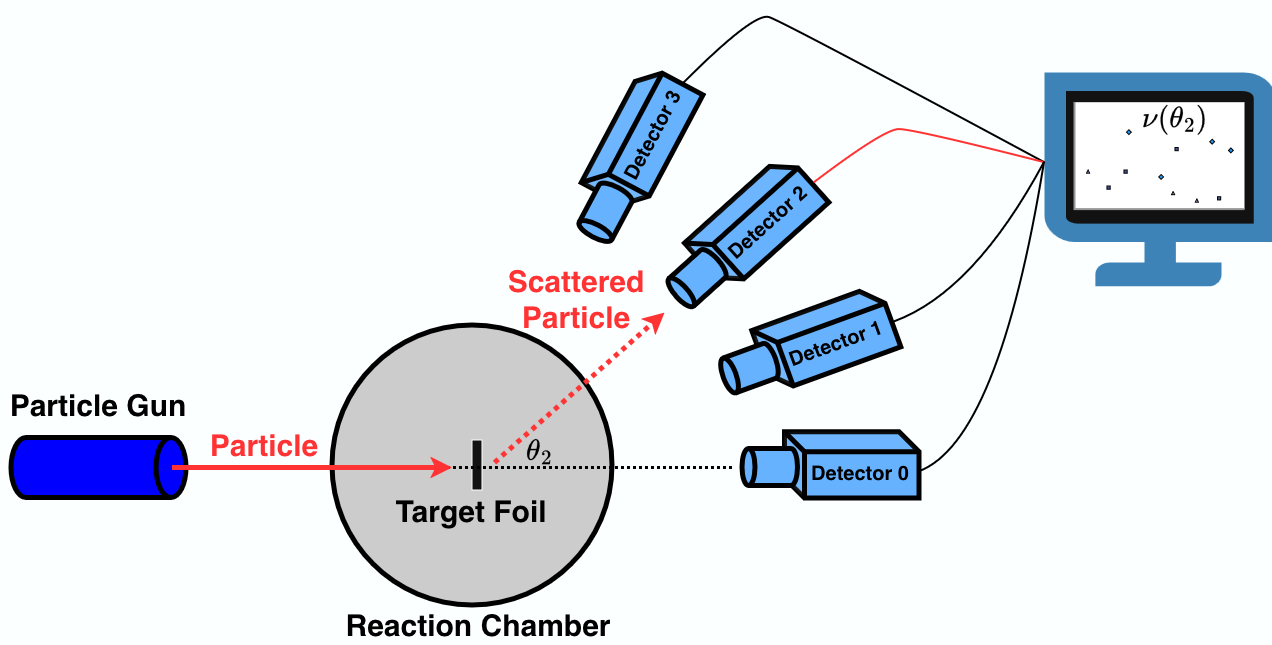}
    \caption{Schematic representation of the Rutherford experiment: A particle gun fires a particle on a target foil. The scattered particle will be detected by one of the four detectors that are mounted at fixed angles. Each detector counts the number of particles (per time) that are leaving the reaction chamber at that specific angle. The scenario presented here displays a particle being scattered at angle $\theta_2$ and registered by detector 2. The observable reported to the experimentalist is the particle rate $\nu(\theta_2)$.}
    \label{fig:fig_3}
\end{figure}

To systematically evaluate the proposed multi-dataset SAGIPS framework, we construct a controlled experiment inspired by the famous Rutherford setup~\cite{NP}. The experimental design, presented in Fig.~\ref{fig:fig_3}, enables us to assess how the framework performs as the number of datasets and/or GPUs varies and how computing resources are utilized. The experiment provides a clear benchmark for validating the quality and stability of the recovered unknowns.

The setup presented in Fig.~\ref{fig:fig_3} shows how different detectors measure the particle rate $\nu(\theta_i)$, i.e. how many particles per time interval are registered at a given scattering angle $\theta_i$. The particle rate can be computed via the Rutherford scattering formula~\cite{NP}:
\begin{equation}
    \label{def_nu_ruth}
    \nu(\theta_i) = C_{exp}(i) \cdot \exp(2a_{0}) \cdot \frac{\sin^{a_1}(\theta_i/2)}{(4T)^2} \cdot [1+\delta_{\nu}(i)\cdot \mathcal{N}(0,1)].
\end{equation}

The constant $C_{exp}(i)$ captures  experimental settings such as the efficiency of the detector (the higher the better) or the particle charge number. The particles kinetic energy (i.e. how fast it is traveling) before hitting the target is expressed as $T$. The uncertainty of each experiment, e.g. measuring dead time or mis-calibration, is expressed by $\delta_{\nu}(i)$. The parameters $a_0 = \ln\Big(\frac{197.3}{137}\Big)\approx0.365$ and $a_1=-4$ have been established in nuclear physics experiments and are thus known very well in literature~\cite{NP}. For this work however, we pretend to have no knowledge about $(a_0,a_1)$ and therefore treat them as the unknowns we wish to extract with our multi-dataset framework. It should be noted that units of all variables in Eq.~\ref{def_nu_ruth} have been omitted, since this setup represents a toy problem and not a real physics measurement. Furthermore, $\nu(\theta_i)$ is expressed in a form that is mathematically equivalent to the standard textbook expression but has been adapted to better suit our experimental design.
 
\subsection{Loop Closure Tests and Datasets}
\label{sec:exp_data}
The performance of SAGIPS is assessed through loop closure tests, for which we create toy datasets $R_i$ with known pipelines $P_i$:
\begin{equation}
\label{defF_exp}
 R_i = P_i[a_0,a_1,\theta_i,T,C_{exp}(i),\delta_{\nu}(i)] = \ln(\nu(\theta_i))
\end{equation}
The natural logarithm on $\nu$ is used for convenience, i.e. the products and fractions in Eq.~\ref{def_nu_ruth} become summations and differences. The pipeline made available to SAGIPS is structurally similar:
\begin{equation}
    \label{defB_exp}
    B_i = P_i[A_0,A_1,\theta_i,T,C_{exp}(i),0.1]
\end{equation}
with $A_0$ and $A_1$ replacing $a_0$, $a_1$ in Eq.~\ref{defF_exp} and representing the generator predictions. Since the inputs $a_0$ and $a_1$ are clearly defined, we can compare them directly to $A_0$ and $A_1$, providing clean and unambiguous performance metrics.

A quick comparison between Eq.~\ref{defF_exp} and ~\ref{defB_exp} reveals that the latter has a fixed uncertainty $\delta_{\nu}=0.1$ while the former does not. This discrepancy is intentional and reflects a realistic experimental scenario: while the forward process $P_i$ is assumed to be well understood, uncertainties in the measurement process — such as misaligned or faulty equipment — or sub-processes that are not fully accounted for may introduce small variations in $R_i$ that are not captured in $P_i$. It is therefore important to take all datasets into account, because (a) they all carry the unknowns $a_0$,...$a_{M-1}$ and (b) no single dataset can be assumed to be definitively more informative than the others without prior knowledge. 

\begin{figure}[htbp]
    \centering
    \includegraphics[width=0.47\textwidth,origin=l]{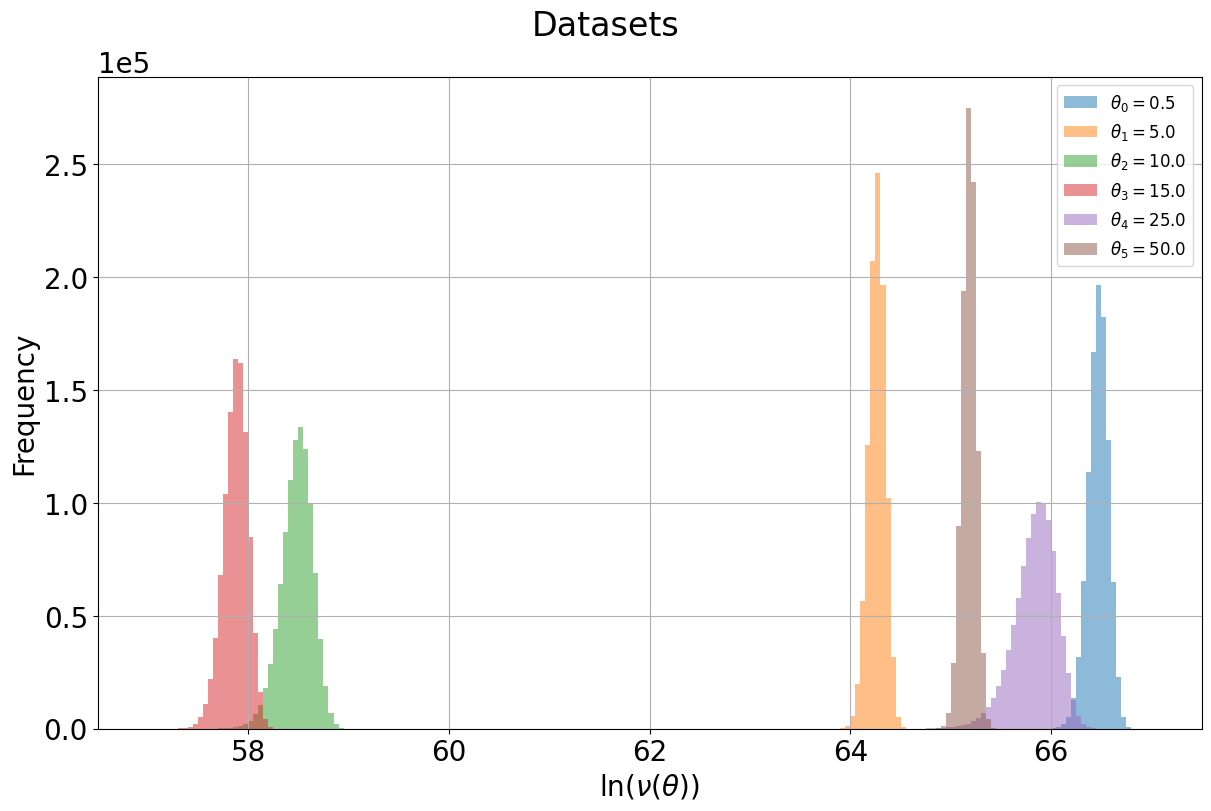}
    \includegraphics[width=0.46\textwidth,origin=r]{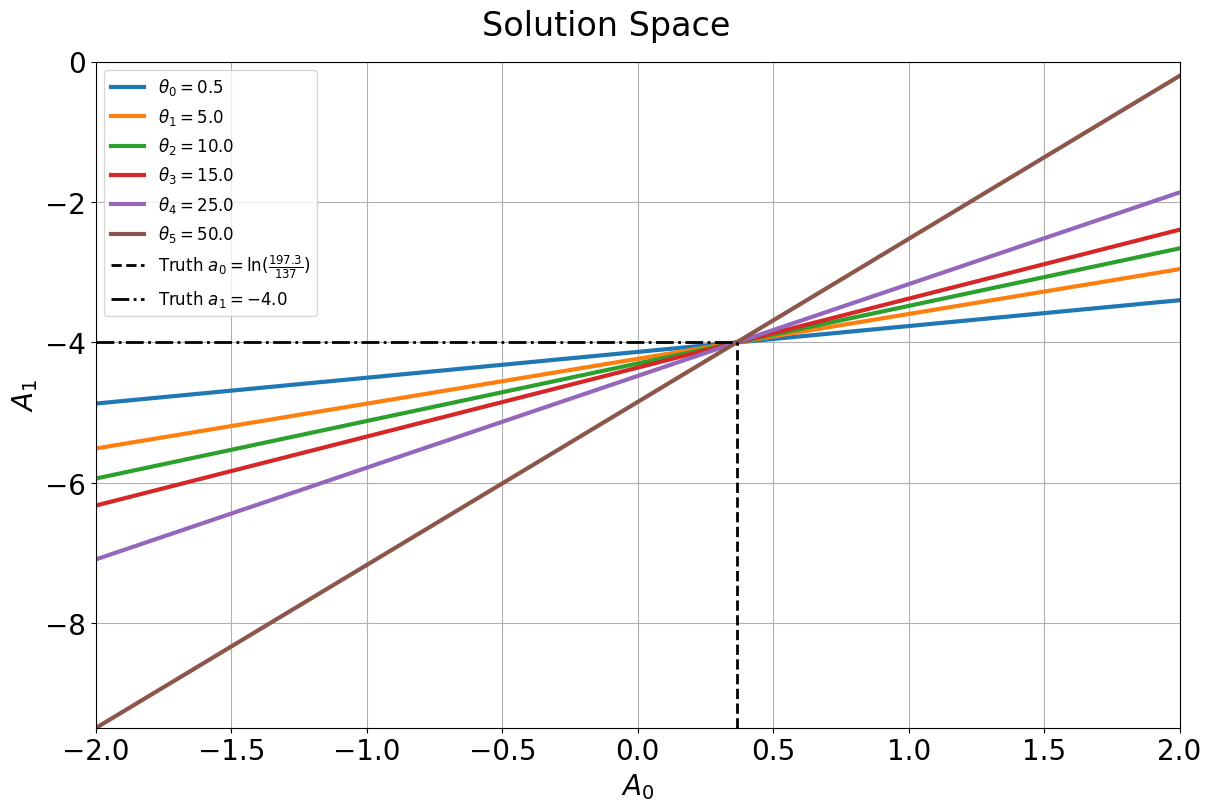}
    \caption{{\bf Left:} Datasets $R_i$, created via Eq.~\ref{defF_exp} in combination with the settings in Tab.~\ref{tab:tab_1}. {\bf Right:} Solution space for each dataset $R_i$. For visualization purposes, the $\delta_v(i)$-term has been omitted for each plot. The dashed horizontal and dashed-dot vertical lines represent the true values $a_0$ and $a_1$ respectively.}
    \label{fig:fig_data}
\end{figure}

Using Eq.~\ref{defF_exp} together with the settings presented in Tab.~\ref{tab:tab_1}, we create six datasets, each consisting of $10^6$ numerical values for $\ln(\nu)$. The left panel in Fig.~\ref{fig:fig_data} displays all six datasets, each effectively covering a different region of the $\ln(\nu)$ feature space. While the values of $\ln(\nu)$ are all of similar order of magnitude $\mathcal{O}(10)$, the distributions differ in their center values and spreads. 
\begin{table}[htbp]
    \centering
    \begin{tabular}{c||c|c|c|c}
     Dataset $R_i$ & $\theta_i$ [Deg] & $\ln(C_{exp}(i))$ & $T$ [a.u.] & $\delta_{\nu}(i)$ \\
     \hline
     \hline
      $R_0$ & $0.5$ & $50.0$ & $5.0$ & $0.1$ \\
      $R_1$ & $5.0$ & $57.0$ & $5.0$ & $0.08$ \\
      $R_2$ & $10.0$ & $54.0$ & $5.0$ & $0.15$ \\
      $R_3$ & $15.0$ & $55.0$ & $5.0$ & $0.12$ \\
      $R_4$ & $25.0$ & $65.0$ & $5.0$ & $0.2$ \\
      $R_5$ & $50.0$ & $67.0$ & $5.0$ & $0.07$
    \end{tabular}
    \caption{Settings to create 6 datasets $R_i$, following Eq.~\ref{defF_exp}. The dataset indices $R_i$ presented in the left column reflect the order in which the datasets are analyzed by SAGIPS.}
    \label{tab:tab_1}
\end{table}
The solution space $(A_0,A_1)$ --- which we wish to explore with the GAN --- is presented in the right panel of Fig.~\ref{fig:fig_data} and follows a linear relationship between $A_0$ and $A_1$. The points on each line represent a possible pair $(A_0,A_1)$ that satisfies Eq.~\ref{defF_exp} in the absence of measurement uncertainties. All lines intersect in one specific point which satisfies: $A_0=a_0$ and $A_1=a_1$. This demonstrates that at least two datasets must be analyzed simultaneously to uniquely determine the two unknowns $a_0$ and $a_1$. Analyzing a single dataset would therefore yield an underdetermined system with infinitely many solutions.

While the extraction of two scalar unknowns from Rutherford scattering is analytically tractable and would not require a GAN-based approach in practice, this setup is deliberately chosen for its simplicity and interpretability — the known analytical solution provides an unambiguous benchmark for validating the proposed framework, and the two-dimensional solution space $(A_0, A_1)$ allows direct visualization of the method's convergence behavior. 

\subsection{SAGIPS Setup}
\label{sec:exp_setup}
Experiments are conducted by incrementally increasing the number of datasets provided to SAGIPS, starting with one and progressing to all six. The order in which the datasets are read in successively is given in Tab.~\ref{tab:tab_1}, i.e. we analyze dataset $R_0$, then datasets $R_0,R_1$, then datasets $R_0,R_1,R_2$ and so on. We utilize an ensemble of independently initialized GANs~\cite{lakshminarayanan2017simple} to enable uncertainty quantification, with each experiment beginning from a freshly initialized ensemble to avoid contamination from previous runs. We did not fix seeds for ensemble members across experiments. Consequently, results from different methods or configurations reflect independent stochastic realizations rather than matched initializations. The generator and discriminator architectures follow those established in prior SAGIPS work~\cite{SAGIPS_2025}: four hidden layers with 128 neurons each and leaky‑ReLU activation functions. Training employs an ensemble of $10$ GANs, each trained for $10\,\rm{k}$ epochs. The generator predicts $N_G=1\,\rm{k}$ samples, and for each generated prediction, the pipeline $P_i$ generates $100$ synthetic realizations $B_i$. The discriminator therefore processes: 
\begin{equation}
    \label{def_ND}
    N_D = 100 \cdot N_G
\end{equation}
synthetic events per update, compared against an equal number of real events drawn from the corresponding dataset $R_i$. The learning rates are set to $10^{-4}$ for the discriminator and $10^{-5}$ for the generator, using Adam optimization. Hyperparameters such as ensemble size, epoch count, batch sizes, and learning rates, were determined through manual tuning, guided by loss convergence behavior and the quality of the resulting predictions relative to the known ground truth, a validation strategy made possible by the availability of ground truth in this controlled setting. The optimization problem of Eqs.~\ref{def_sagips_gan1} and~\ref{def_sagips_gan2} can be challenging to solve and regularization is employed commonly for better stability during the GAN training process~\cite{SalimansGoodfellowZarembaEtAl16, ArjovskyChintalaBottou17, GulrajaniAhmedArjovskyEtAl17, RothLucchiNowozinEtAl17, ZhangGoodfellowMetaxasEtAl19}. In our experiments, however, the training does not suffer instabilities and does not require any regularization. The simplicity and low dimensionality of the Rutherford benchmark, combined with large synthetic dataset sizes, reduce typical GAN instabilities. We perform two categories of runs:
\begin{itemize}
    \item {\bf Run Group A:} Experiments that are either carried out on a single GPU or on multiple GPUs with multiplicity $m=1$, i.e. we strictly use one GPU per dataset.

    \item {\bf Run Group B:} Experiments are carried out on multiple GPUs with a multiplicity $m>1$, i.e. each of the $N$ subsets is divided across $m$ GPUs.
\end{itemize}
This division allows us to isolate the effects of dataset count and multiplicity.

\subsection{Single GPU Runs}
\label{sec:exp_single_gpu}
In single‑GPU experiments, the full set of datasets is either concatenated and passed through a single discriminator (single discriminator setting) or assigned individual discriminators (multi-discriminator setting) whose gradients are combined before updating the generator. In both settings, the generator produces a single set of predictions that is passed through each dataset‑specific forward operator. The resulting synthetic datasets are either merged and analyzed by a single discriminator or evaluated separately by their corresponding discriminators. The multi-discriminator setting mimics the conceptual structure of our distributed approach but without inter‑GPU communication. The single‑GPU analyses are computationally feasible only in this synthetic example (and other problems of similar complexity). Real‑world applications of SAGIPS typically exceed single‑GPU memory and compute limits.

\subsection{Multi-GPU Runs}
\label{sec:exp_multi_gpu}
In the multi‑GPU setting, each dataset is assigned to its own GPU, enabling parallel evaluation of dataset‑specific forward operators and discriminators. This configuration mirrors the conceptual architecture described in Section~\ref{sec:theory_dmda}. Since the proposed multi-dataset framework relies on gradient-transport mechanisms to synchronize training across replicas, the resulting scaling behavior and precision could, in principle, depend on the specific transport mechanism employed. To assess whether our conclusions are robust to this choice, and to understand how the two-level gradient aggregation — across datasets, and via sharding within each dataset — affects transport performance, we evaluate multiple gradient-transport strategies, some of which were introduced in~\cite{SAGIPS_2025}.

\subsubsection{Gradient Transport}
\label{sec:exp_grad_transport}
All gradient-transport methods are implemented using \texttt{mpi4py}~\cite{mpi4py} for inter-GPU communication. We assess four such methods:
\begin{itemize}
    \item \textbf{Conventional ARAR:} The conventional asynchronous ring-allreduce utilizes all available GPUs to accumulate generator gradients, providing maximal cross-GPU communication but heavily relying on inter-node communication speed for performance.

    \item \textbf{ARAR:} This ring-allreduce approach uses a grouping mechanism such that GPUs located on the same node (the inner group) communicate more frequently than GPUs on different nodes (the outer group). Inter-node communication is further restricted by limiting the outer group to one GPU per node. For example, an analysis with $10$ nodes and four GPUs each has an inner group size of four and an outer group size of $10$. This approach has been shown to be faster than its conventional counterpart~\cite{SAGIPS_2025}. In the analyses conducted here, the outer group communication occurs every 10-th training epoch.

    \item \textbf{Strong ARAR:} A hybrid between the two previously discussed methods. Similar to ARAR, inner group communication occurs more frequently than outer group communication. However, the outer group now includes all available GPUs, combining the lower communication cost of ARAR with the broader synchronization of conventional ARAR. The outer group update frequency was also set to $10$ epochs.

    \item \textbf{Double Binary Tree:} A recently added method to SAGIPS following the work presented in~\cite{binarytree2013} and~\cite{SANDERS2009581}. The power-of-two constraint for ordering GPUs is not enforced in this implementation.
\end{itemize}

\subsubsection{Impact of Multiplicity}
\label{sec:exp_multiplicity}
The number of GPUs assigned to each dataset is denoted as the multiplicity $m$. When datasets are sharded ($m>1$), the number $N_G$ of samples predicted by the generator scales inversely with $m$ according to:
\begin{equation}
    \label{def_NG}
    N_G = \Big\lfloor \frac{1\,\rm{k}}{m} \Big\rfloor,
\end{equation}
where $1\,\rm{k}$ is the default number of generated samples used when data is not sharded. Increasing $m$ therefore reduces discriminator throughput per step, but also decreases the memory and compute demands placed on each GPU. This experiment probes whether higher multiplicities accelerate training without degrading inference quality.

\subsection{Computational Resources}
\label{sec:exp_computing}
All experiments from run group A were conducted on the Jefferson Lab computing farm, using a single node with $8$ NVIDIA A800 GPUs. Experiments for run group B were carried out on the Polaris HPC system\footnote{\url{https://www.alcf.anl.gov/polaris}}, hosted at Argonne National Lab. Each Polaris node is equipped with 4 NVIDIA A100 GPUs, one AMD EPYC "Milan" processor and a HPE slingshot 11 network. 

\section{Evaluation and Results}
\label{sec:eval}
We evaluate the performance of the multi-dataset SAGIPS framework by analyzing the accuracy of the recovered unknowns, the stability of the ensemble predictions, and the computational characteristics of the training process. Following the evaluation procedure established in~\cite{SAGIPS_2025}, all generator predictions reported in this work are obtained by averaging across all rank replicas within each ensemble member before computing the ensemble mean. This rank averaging is applied consistently across all gradient transport methods and serves as the primary mechanism for mitigating the impact of rank drift on the final parameter estimates. The primary metric is the relative residual:
\begin{equation}
   \label{def_r}
    \hat{r}_i = \frac{a_i - A_i}{a_i}
\end{equation}
where $a_i$ is the true value of the unknown and $A_i$ the ensemble-averaged prediction. The associated uncertainty is computed from the ensemble standard deviation $\Delta A_i$:
\begin{equation}
    \label{def_dr}
    \Delta\hat{r}_i = \frac{\Delta A_i}{|a_i|}
\end{equation}
Training time is determined via an offline analysis of timestamped checkpoints — each GAN in the ensemble is written to disk at regular intervals, yielding 20 checkpoints in total. The training time for a given checkpoint is defined as the time at which every GPU, across all ensemble members, has finished writing its copy of the generator to disk. The total training time is then given by the difference between the timestamps of the checkpoints at the beginning and end of training. GPU compute and memory utilization are monitored via the \texttt{pyNVML}\footnote{\url{https://pypi.org/project/nvidia-ml-py/}} package. All reported GPU utilization values represent averages across all ranks and ensemble members.

\subsection{Results from Run Group A}
\label{sec:eval_A}

\subsubsection{{Convergence Quality}}
\label{sec:eval_A_conv}
\begin{figure}[htbp]
 \centering
 \includegraphics[width=0.95\textwidth]{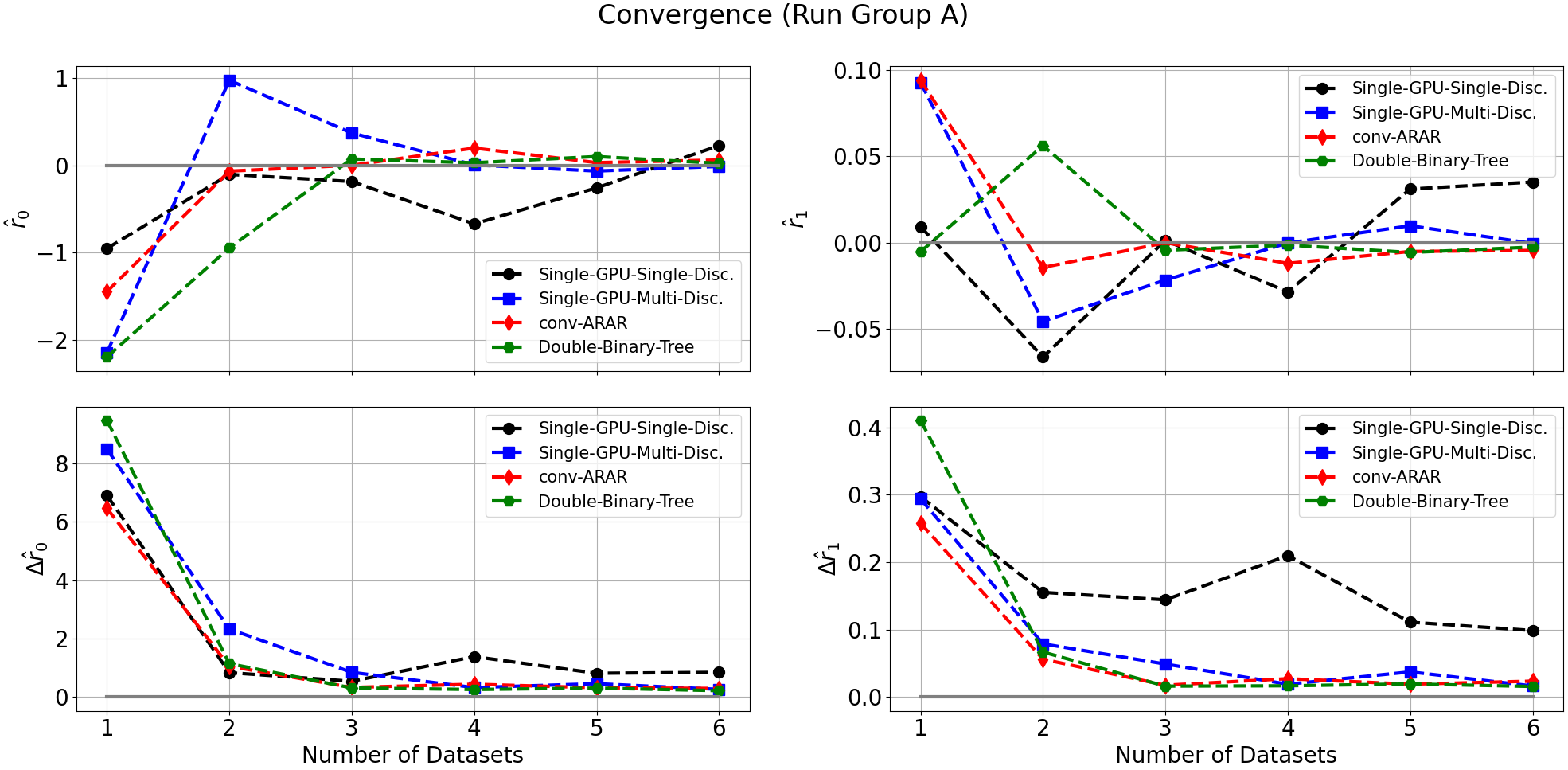}
 \caption{{\bf Top Row:} Relative residual from Eq.~\ref{def_r} for unknown $a_0$ (left) and $a_1$ (right) as a function of the number of datasets fed into SAGIPS. {\bf {Bottom Row:}} Uncertainties associated with $\hat{r}_0$ (left) and $\hat{r}_1$ (right) as a function of the number of analyzed datasets. The different curves in each panel present the various analysis approaches explored within this run group.}
 \label{fig:compare_ifarm}
\end{figure}

Figure~\ref{fig:compare_ifarm} shows the relative residuals and their associated uncertainties as a function of the number of datasets analyzed. The order in which the datasets are added to the analysis follows the sequence (top to bottom) laid out in the left column in Tab.~\ref{tab:tab_1}. A brief discussion on how altering that order might affect the performance follows in a later section. 
All approaches summarized in Fig.~\ref{fig:compare_ifarm} produce residuals that are, within the given uncertainties, consistent with zero and thus reconstruct the unknowns $(a_0, a_1)$. However, the accuracy of the predictions is poor for all methods when only one dataset is available. This is expected because a single dataset defines a one-dimensional manifold in the solution space $(A_0, A_1)$, which is insufficient to simultaneously constrain two unknowns (see right panel of Fig.~\ref{fig:fig_data}). As additional datasets are incorporated, the residuals approach zero and their uncertainties shrink accordingly. The single-GPU single-discriminator baseline exhibits the weakest performance, particularly for $\Delta\hat{r}_1$. A single discriminator might not be able to properly handle the varying feature ranges of the individual datasets. The single-GPU multi-discriminator and Double-Binary-Tree configurations show, compared to the other methods, a pronounced deviation from zero for both residuals at $n=2$. We attribute this observation to the fact that solutions produced at the $n=2$ configuration are more sensitive to the properties (e.g. noise level) of each dataset and to the optimization trajectory of the individual solver. Unlike $n>2$, where additional datasets provide redundant constraints that help averaging out these inconsistencies, the $n=2$ configuration offers no such redundancy, because the two datasets provide only the minimum information needed to determine the two unknown parameters. A more detailed analysis, including fixed seeds across the compared runs, would help isolating the driving mechanism behind the $n=2$ sensitivity, e.g. variance from random initialization vs. variance introduced by the underlying gradient transport mechanism. Since this would go beyond the scope of this work, we leave this task open for future studies. Aside from the $n=2$ sensitivity, the multi-GPU configurations (conv-ARAR, Double-Binary-Tree) achieve the strongest and most consistent recovery overall. This suggests that (a) the multi-dataset analysis method is well suited for extracting unknowns from multiple data sources simultaneously, and (b) the underlying gradient transport method has no noticeable impact on the ability of the approach to recover unknowns consistent with the ground truth.

\subsubsection{Monitored GPU Resource Utilization}
\label{sec:eval_A_resource}
\begin{figure}[htbp]
    \centering
    \includegraphics[width=0.5\textwidth]{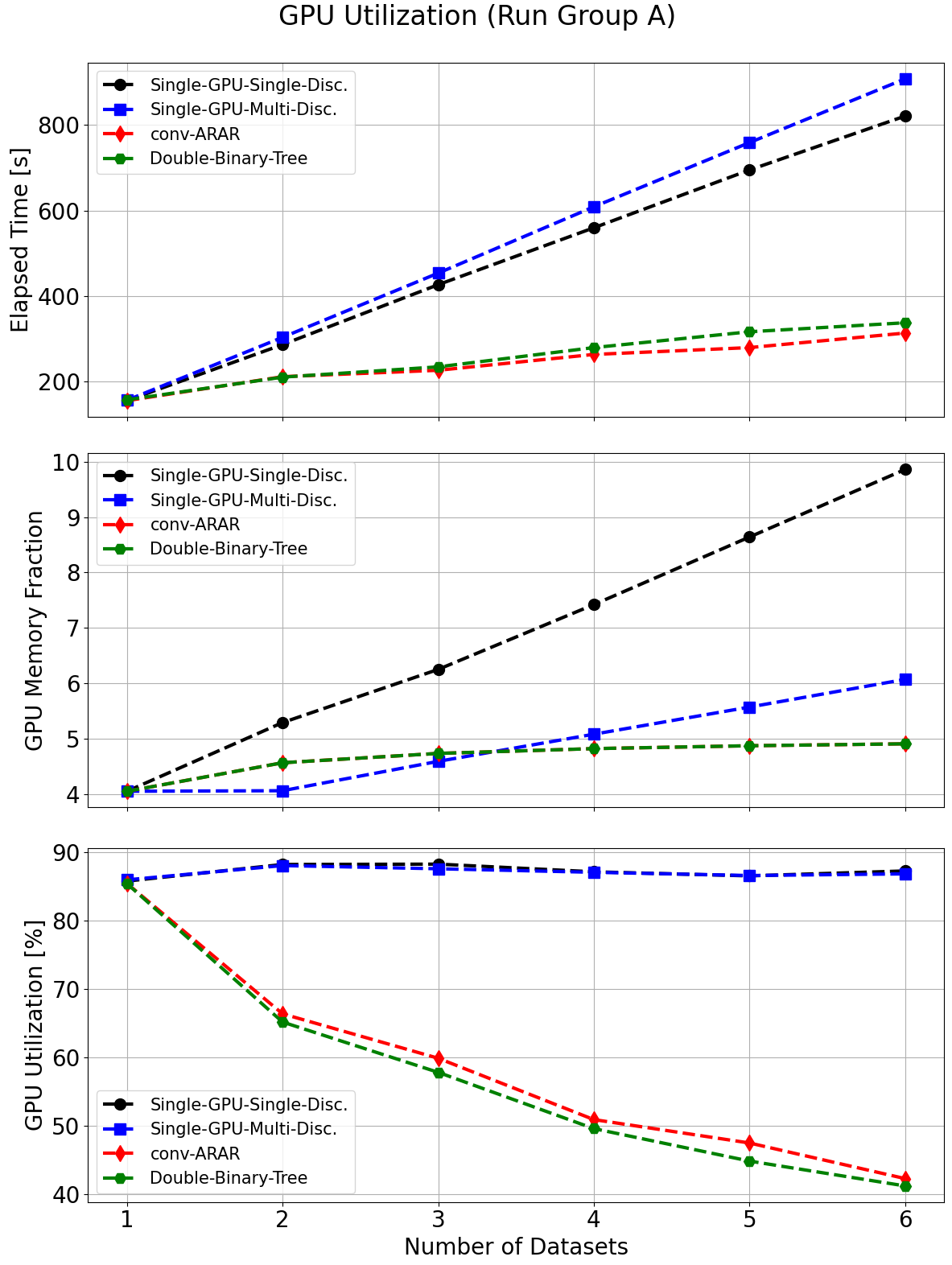}
    \caption{Total training ({\bf top panel}), GPU memory fraction ({\bf center panel}) and GPU utilization ({\bf bottom panel}) as a function of the number of datasets used for the SAGIPS analysis.}
    \label{fig:summary_ifarm}
\end{figure}

Figure~\ref{fig:summary_ifarm} summarizes the total training time, GPU memory fraction, and GPU utilization. The top panel of Fig.~\ref{fig:summary_ifarm} shows that the single-GPU multi-discriminator approach requires the longest training time. 
This is expected, because multiple discriminator models are trained within a single SAGIPS update on one device. The single-GPU single-discriminator setup is approximately $100\,\rm{s}$ faster due to its simpler model structure. Both multi-GPU approaches achieve a factor of $\gtrsim 2$ improvement in training time through distributed computation. The center panel shows that the single-GPU single-discriminator approach requires the most GPU memory, since all datasets must reside simultaneously in the memory of a single A800 GPU. A similar memory trend holds for the single‑GPU multi‑discriminator setup, though the data is distributed across multiple models rather than a single one. The multi-GPU configurations show near-constant memory requirements per GPU, independent of the total number of datasets analyzed. This behavior follows directly from the distributed design — each GPU is responsible for processing a single dataset regardless of how many datasets are included in the analysis, resulting in an approximately fixed memory footprint per GPU. The bottom panel shows that GPU utilization decreases as more GPUs are added in the multi-GPU configurations. This behavior is attributed to the CPU-based gradient aggregation mechanism: each algorithm offloads local gradients to CPU memory before distributing them via \texttt{mpi4py}. Once accumulated and averaged, the gradients are reloaded into GPU memory and used for updating the weights of the local generator copy. As the number of GPUs increases, these CPU-bound operations become increasingly dominant, reducing average GPU utilization. This is not a design flaw but an inherent consequence of the communication patterns implemented in this particular workflow~\cite{SAGIPS_2025}. The single-GPU configurations show approximately constant utilization independent of the number of datasets, consistent with their fixed computational structure. 

\subsubsection{Assessing possible systematic Bias from the Input Data}
\label{sec:eval_A_order}
\begin{figure}[htbp]
    \centering
    \includegraphics[width=0.85\textwidth]{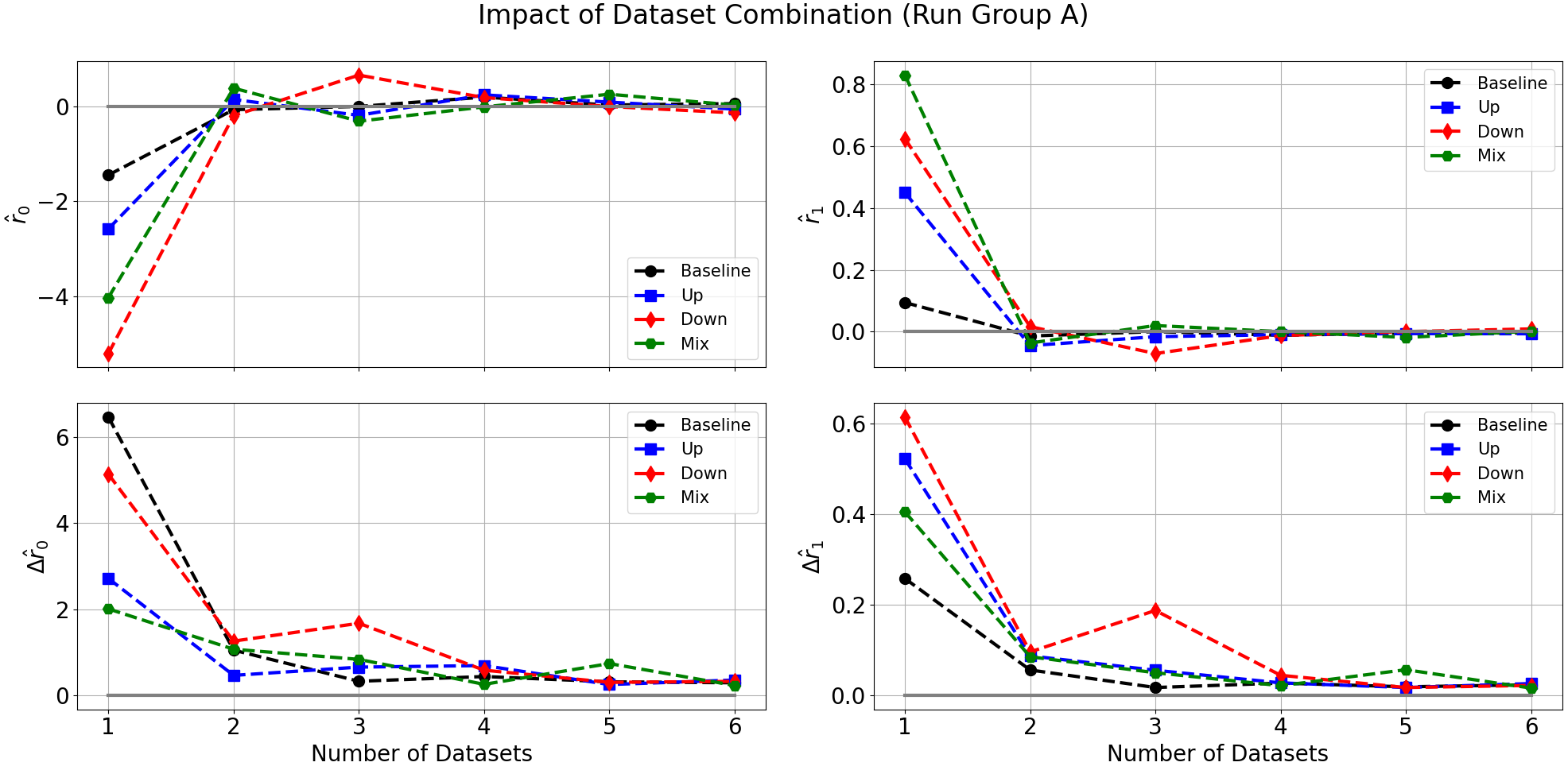}
    \caption{{\bf Top Row:} Relative residual from Eq.~\ref{def_r} for unknown $a_0$ (left) and $a_1$ (right) as a function of the number of datasets fed into SAGIPS. {\bf {Bottom Row:}} Uncertainties associated with $\hat{r}_0$ (left) and $\hat{r}_1$ (right) as a function of the number of analyzed datasets. The different curves in each panel present the different ordering schemes (explained in main text) for successively combining and analyzing the data. Each analysis was carried out with the conventional ARAR gradient transport.}
    \label{fig:compare_ifarm_order}
\end{figure}

An important question that may arise at this point is whether, and how, the data that is fed into the inverse solver might affect its performance. For example, a dataset with very poor experimental resolution might degrade the precision of the extracted unknowns. One way to assess such potential impact is to follow the procedures laid out in this work, namely feeding one, two, three, ... datasets into the inverse solver and run multiple passes in which the order of combination is varied. Because the datasets are combined sequentially, each intermediate point ($n<6$) corresponds to a scenario in which only a specific subset of datasets is available, while the remainder are effectively excluded. Varying the ordering scheme in a specific manner allows us to probe the impact of dataset exclusion. Fig.~\ref{fig:compare_ifarm_order} summarizes the results of this particular study.
We designed three ordering schemes for combining the six datasets that are analyzed by SAGIPS, using the conventional ARAR gradient transport. The schemes are defined by the dataset-specific resolution $\delta_\nu(i)$, as listed in Tab.~\ref{tab:tab_1}. The first scheme, \textit{Up}, combines datasets in ascending order of $\delta_\nu$, so that for $n<4$ it corresponds to a scenario where only the highest-resolution datasets are available. The second scheme, \textit{Down}, combines datasets in descending order of $\delta_\nu(i)$, so that for $n<4$ it corresponds to the opposite case, where only the lowest-resolution datasets are available. The third scheme, \textit{Mix}, is a random permutation of the default dataset order given in Tab.~\ref{tab:tab_1}. The residuals and their uncertainties are recorded for each scheme and compared to the baseline analysis (black curves in Fig.~\ref{fig:compare_ifarm_order}). As expected, all schemes converge to the same performance once all six datasets are analyzed, since every configuration then has access to identical information. Among all configurations and runs, the \textit{Down} scheme stands out at $n=3$, where residual quality visibly worsens for both unknowns. This configuration corresponds to a scenario in which only the three lowest-resolution datasets are analyzed. The resulting residuals remain statistically consistent with zero, but at the cost of substantially larger uncertainty (red curves, bottom row). Performance recovers markedly once a fourth dataset is added, suggesting that the low-resolution subset alone is comparatively poorly constrained, and that adding higher-resolution data disproportionately improves precision. The \textit{Up} and \textit{Mix} schemes, by contrast, remain consistent with the baseline within uncertainties across all probed points $n$. These results demonstrate that the specific subset of datasets included (or left out) in a multi-dataset analysis can materially affect performance, particularly when that subset happens to consist predominantly of low-resolution data. As a systematic check, we therefore recommend running multi-dataset analyses under multiple dataset orderings or configurations to assess the stability of the obtained results against the possibility of missing or excluded datasets.

\subsection{Results from Run Group B}
\label{sec:eval_B}
The Run Group B experiments were carried out on the Polaris machine with multiplicity $m=20$, i.e. each dataset was split across 20 GPUs. We followed once again the order displayed in the left column of Tab.~\ref{tab:tab_1} to feed the datasets into the inverse solver. In order to keep node-hour utilization at a reasonable level, we conduct the multi-GPU studies for analyzing all 6 datasets and only run a full sweep over all dataset configurations for the conventional ARAR gradient transport.
\begin{figure}[htbp]
    \centering
    \includegraphics[width=0.85\textwidth]{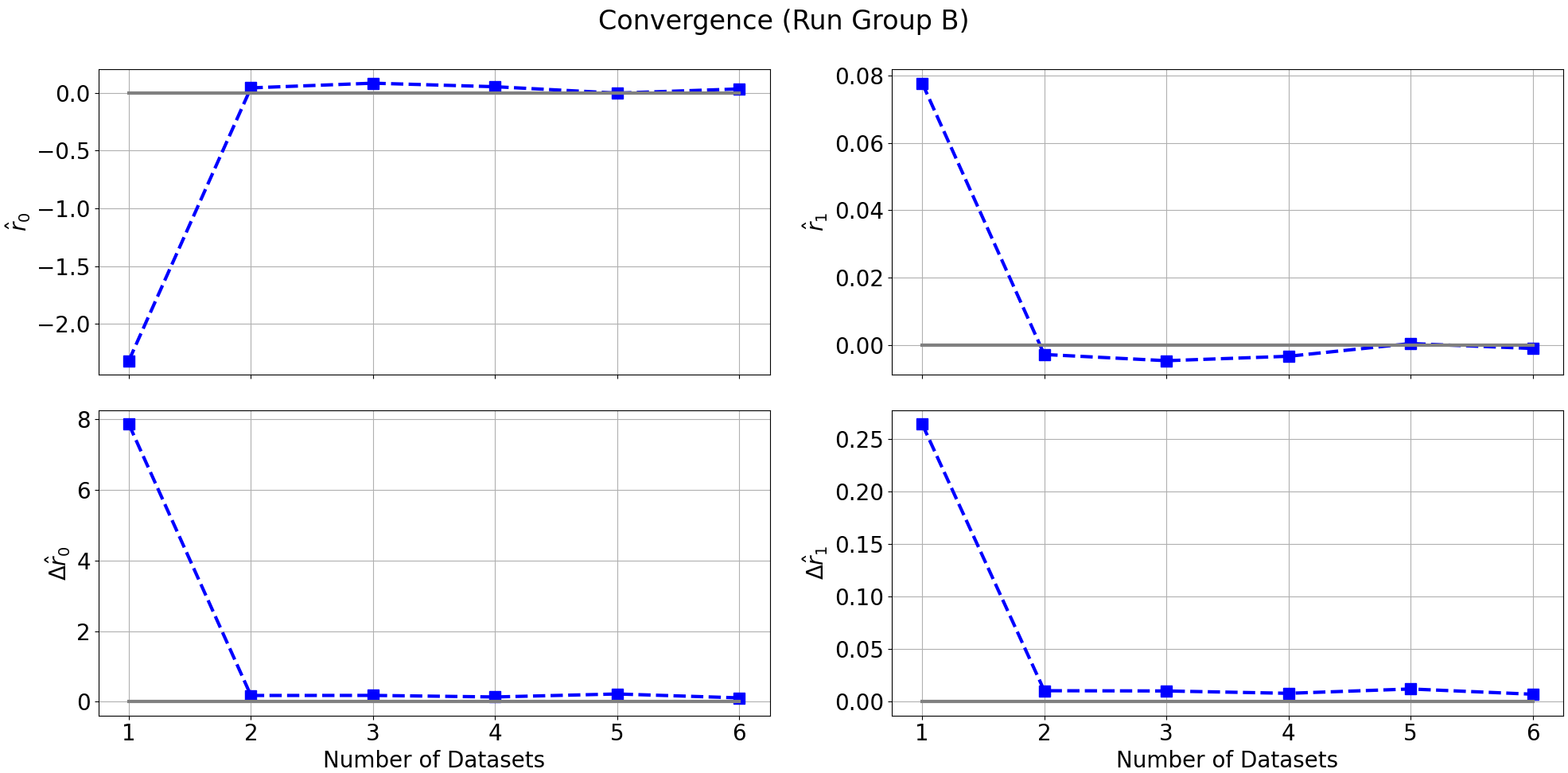}
    \caption{{\bf Top Row:} Relative residual from Eq.~\ref{def_r} for unknown $a_0$ (left) and $a_1$ (right) as a function of the number of datasets fed into SAGIPS. {\bf {Bottom Row:}} Uncertainties associated with $\hat{r}_0$ (left) and $\hat{r}_1$ (right) as a function of the number of analyzed datasets. All curves shown here have been obtained by using the conventional ARAR gradient transport method.}
    \label{fig:fig_conv_arar_polaris}
\end{figure}
Fig.~\ref{fig:fig_conv_arar_polaris} presents the normalized residuals $\hat{r}_0$ and $\hat{r}_1$, as well as their corresponding uncertainties, as a function of the number of analyzed datasets for the conventional ARAR gradient transport. The performance is poor (i.e. residuals in-consistent with zero) when only one dataset is analyzed, but improves noticeably as soon as more datasets are added to the analysis. These results are consistent with the observations made in section~\ref{sec:eval_A} and indicate that the additional data sharding, in the context of multi-dataset analyses, has no negative impact on the inverse solver's performance. 

Tab.~\ref{tab:tab_polaris_summary} summarizes the normalized residuals, training times and GPU utilization for all four gradient transport methods. The underlying analyses had access to all six datasets. The GPU memory utilization is not displayed because it turned out to be a constant value of $3.47$ for all methods. The normalized residuals are statistically consistent with zero, independent of the gradient transport method. However, there are differences in the convergence quality which shall be briefly discussed in the following. The grouped ARAR method shows inferior accuracy. While reducing the inter-node communication accelerates training, it restricts gradient sharing and reduces the generator’s ability to reconcile all dataset constraints. In contrast, conventional ARAR and the double binary‑tree method require training times in the order of multiple hours, but achieve the best parameter reconstruction, confirming that rich cross‑node communication improves global consistency. Strong‑ARAR requires noticeably less training time but performs competitively, offering a balance between communication cost and convergence quality. The GPU utilization is larger for ARAR and strong ARAR, because the gradient transfer is mainly focused on GPUs within a single node. The internode communication happens only every 10 update states which reduces the communication overhead and therefore keeps the individual GPU less idle. 
\begin{table}[htbp]
    \centering
    \renewcommand{\arraystretch}{1.5}
    \begin{tabular}{c||c|c|c|c|}
     & conv-ARAR & Double Binary-Tree & ARAR & strong-ARAR \\
    \hline
    \hline
    $\hat{r}_0 \pm \Delta\hat{r}_0$  & $0.034\pm 0.107$  & $-0.033\pm 0.145$ & $0.697\pm 1.536$ & $0.063 \pm 0.303$ \\ 
    $\hat{r}_1 \pm \Delta\hat{r}_1$  & $-0.001\pm 0.007$  & $-0.003\pm 0.009$ & $-0.033\pm 0.121$ & $-0.0013 \pm 0.024$ \\
    Elapsed Time $[s]$ & $6,334$ & $9,209$ & $2,077$ & $2,243$ \\
    GPU Utilization $[\%]$ & $8.42$ & $1.97$ & $28.73$ & $20.16$ \\
    $\frac{\text{Elapsed Time }(m=4)}{\text{Elapsed Time }(m=20)}$ & $1.521$ & - & - & $3.354$ \\
    \end{tabular}
    \caption{Performance evaluation of the multi-dataset analyses conducted on the Polaris HPC, using a $m=20$ multiplicity. The underlying analyses ran with all six datasets and different gradient transport methods. The last row represents the ratio between training times using a $m=4$ and $m=20$ multiplicity respectively. The ratio was determined for two gradient transport methods only.}
    \label{tab:tab_polaris_summary}
\end{table}

\subsubsection{Scaling with Data Multiplicity}
\label{sec:eval_B_scaling}
Another interesting question in the context of multi-dataset analysis is whether the choice of multiplicity, i.e. the degree of data sharding, has an impact on performance and training time. We therefore determined the training for the conventional ARAR baseline, using two folds $m=4$ and $m=20$ respectively. The ratio between these two times defines the speedup gained by increasing the multiplicity. The bottom row of Tab.~\ref{tab:tab_polaris_summary} lists said ratio for conventional and strong ARAR. We added the ratio of the latter because it has been proven to be a good balance between accuracy and training time. We observe, consistent with previous results, that strong ARAR provides a larger speedup. The quality of the residuals is not noticeably affected when going from $m=4$ to $m=20$ multiplicity, for neither of the two transport methods. Detailed scaling and performance studies for the individual gradient transport methods are left open for future work. 

\subsubsection{Model Drift Analysis}
\label{sec:eval_B_drift}
To assess the impact of the gradient transport method on the multi-dataset analysis we monitor the generator consistency by utilizing two complementary drift metrics. The pairwise weight drift is defined as:
\begin{equation}
\label{def_weight_drift}
d_{ij}^{(w)} = \frac{\|w_i - w_j\|^2}{\frac{1}{N}\sum_{k=0}^{N-1}\|w_k\|^2}
\end{equation}
where $w_i$ denotes the flattened weight vector of the generator replica on rank $i$. The normalized pairwise output drift is defined as:
\begin{equation}
\label{def_output_drift}
d_{ij}^{(o)} = \frac{\|G_{\phi_i}(n) - G_{\phi_j}(n)\|^2}{\frac{1}{N}\sum_{k=0}^{N-1}\|G_{\phi_k}(n)\|^2}
\end{equation}
where $n \sim \mathcal{N}(0, I)$ is the same noise vector fed to all generator replicas. Both metrics are normalized by the mean squared norm across all ranks, making them scale invariant and directly comparable. The output drift metric is particularly valuable in practice as it requires no knowledge of the true parameter values and is therefore applicable to real-world analyses where the ground truth is unavailable.
\begin{figure}[htbp]
    \centering
    \includegraphics[width=0.95\textwidth]{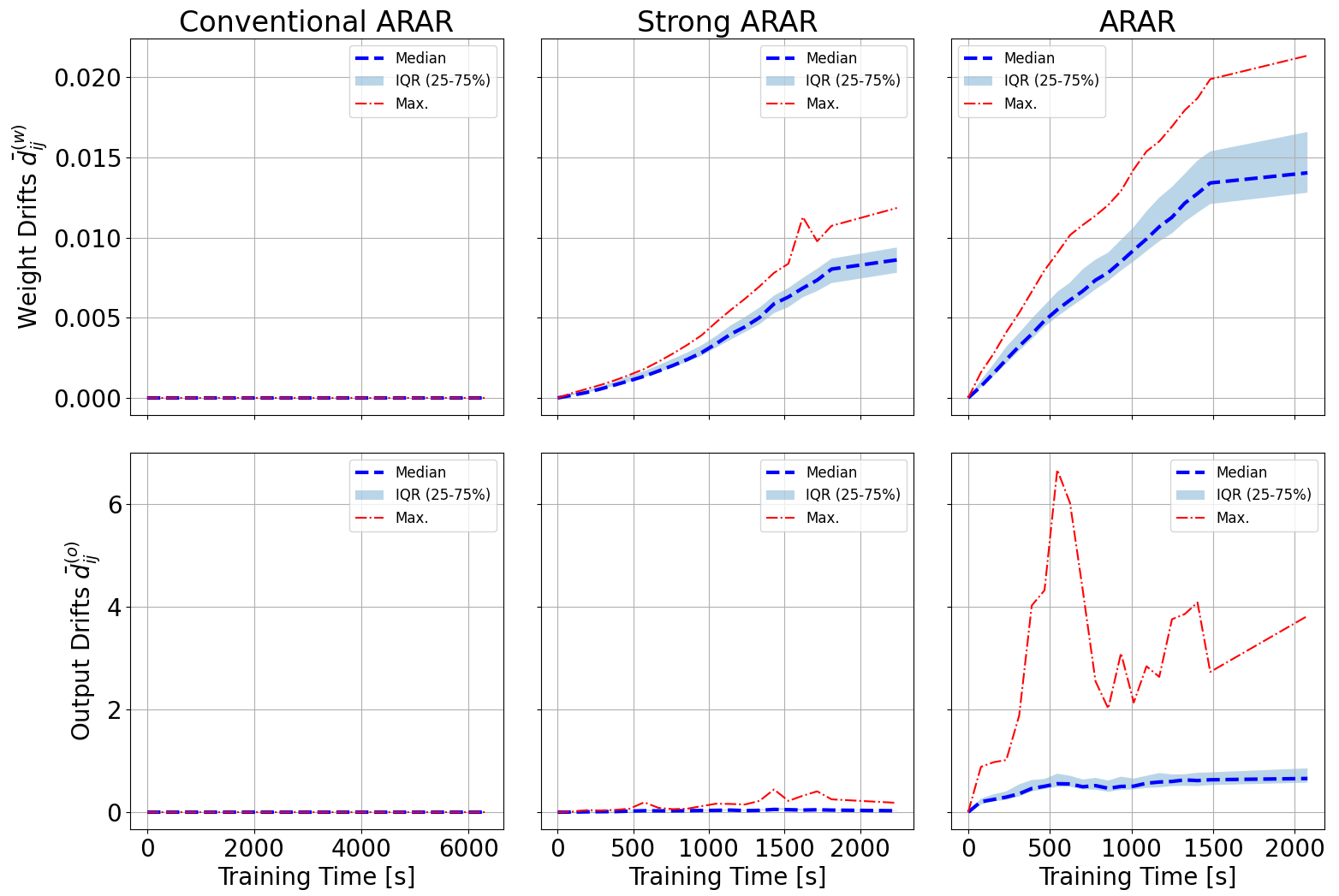}
    \caption{Weight drifts ({\bf top row}) and output drifts ({\bf bottom row}) computed according to Eq.~\ref{def_weight_drift} and~\ref{def_output_drift} as a function of the SAGIPS training times. Shown here are the median, interquantile range (IQR) and maximum determine from the rank pair $(i,j)$ averaged over the ensemble size. The columns (from left to right) represent the pair wise distances for conventional ARAR, strong ARAR and ARAR.}
    \label{fig:compare_drifts}
\end{figure}
Figure~\ref{fig:compare_drifts} summarizes both metrics for conventional ARAR, strong ARAR, and ARAR, where the median interquantile range and maximum across all rank pairs $(i,j)$, and averaged over the ensemble size, are shown. The double binary tree method is not shown as it employs full gradient synchronization across all ranks at every step and is therefore expected to exhibit the same drift behavior as conventional ARAR.

All curves correctly start at zero, reflecting the fact that at the beginning of training all generator replicas within a given ensemble member share identical weights due to the initial synchronization step — the pairwise distance is therefore exactly zero by construction.

Conventional ARAR shows no drift in either metric throughout training, confirming perfect generator consistency across all ranks. Strong ARAR introduces moderate and bounded weight drift, reaching approximately $0.01$ relative units by the end of training, but negligible output drift, demonstrating that its hybrid synchronization strategy effectively suppresses output inconsistency despite moderate weight divergence. ARAR exhibits the largest drift in both metrics, with weight drift growing monotonically throughout training and output drift reaching values significantly above zero for some rank pairs.
These findings are consistent with the convergence quality and training time results presented in Tab.~\ref{tab:tab_polaris_summary}. Conventional ARAR achieves the best parameter recovery but requires the longest training time. Strong ARAR shows equally competitive convergence quality with negligible output drift and a shorter training time. ARAR shows the lowest convergence quality, the largest drift, and the lowest training time, though the training time advantage over strong ARAR is modest. It should be noted that despite the drift observed for grouped methods, all approaches successfully recover the shared unknowns within the reported ensemble uncertainties, as individual rank deviations partially cancel through rank averaging. 

We propose the two drift metrics introduced here as a general diagnostic tool for distributed multi-dataset analyses with a setup similar to SAGIPS. Possible mitigation techniques through improved synchronization strategies or post-training weight averaging is left for future work.

\section{Summary and Outlook}
\label{sec:summary}
In this work, we introduced an extension of the SAGIPS inverse-problem solver that enables the simultaneous analysis of multiple heterogeneous datasets through distributed generative modeling. Each dataset is assigned to a dedicated GPU equipped with its own forward operator and discriminator, while the generator gradients are synchronized across all GPUs via the distributed data parallel formalism. This design addresses the challenges encountered in solving large-scale multi-dataset inverse problems where dataset heterogeneity and computational demands exceed the capabilities of single-device workflows.
Through a series of controlled experiments inspired by Rutherford scattering and conducted on the JLab computing farm as well as the Polaris HPC system, we demonstrated that the proposed approach reliably recovers the underlying physical parameters and that inference quality improves as more datasets are incorporated. The results confirm that treating datasets independently, rather than merging them or training on a single discriminator, produces more accurate and stable parameter estimates. A systematic check varying the order in which datasets are combined confirmed that performance depends not only on the number of datasets analyzed, but also on which specific datasets are available at a given step. Combining only the lowest-resolution datasets transiently degraded precision, though results remained statistically consistent with zero throughout. 
Throughout all experiments, the solver assumed a fixed detector resolution across datasets with differing resolutions. This design mimics a real-world situation where the exact (experimental) setup might be unknown or simply resists precise modeling. Despite this limitation, the quantities extracted by the solver remained consistent with the ground truth, because the solution space was constrained by the diversity and abundance of datasets. A more comprehensive treatment of such systematic effects, including detector-resolution or varying statistical abundance accross datasets, remains an important direction for future work. Our experiments showed furthermore that multi-GPU configurations significantly reduce training time and yield near-constant memory requirements per GPU, extending the applicability of SAGIPS to problems larger and more complex than those discussed in this work. 
This work also showed that further sharding of the data across GPUs is possible and may yield a reduction in overall training time while maintaining the prediction quality that is obtained when no sharding is applied. 
We applied different gradient transport strategies to demonstrate that the underlying assumptions of the multi-dataset approach remain valid across different means to share and accumulate the generator gradients. The generator consistency across ranks was assessed via two complementary drift metrics — pairwise weight drift and normalized pairwise output drift — which revealed differing degrees of synchronization stability across transport methods. Despite this, all methods successfully recovered the shared unknowns within the reported ensemble uncertainties, as individual rank deviations partially cancel in the ensemble mean. The output drift metric, which requires no knowledge of the true parameter values, is proposed as a general diagnostic tool for assessing generator consistency in distributed training settings.
The proposed framework has limitations that practitioners should be aware of. These include, but are not limited to: handling datasets with conflicting physics, systematic biases, adversarial corruptions, insufficient statistical representation, or analyzing datasets that only partially cover the shared unknown space (e.g. where $R_i$ is sensitive to only a subset of $a_0,\ldots,a_{M-1}$). These settings may lead to contradictory or uninformative gradient signals and degraded convergence. In such cases, careful dataset validation, preprocessing, and statistical balancing are recommended prior to applying the proposed framework. We further note that reported results reflect independent stochastic realizations rather than seed-matched comparisons across methods, which should be considered when interpreting small inter-method differences.
A feasibility study conducted on a more complex inverse problem, and summarized in appendix~\ref{sec:app_proxy2d}, clearly demonstrates that the framework generalizes beyond the Rutherford scattering use case. It should be noted that the approach is not limited to GANs — it can be extended to any SAGIPS configuration that uses a trainable model for hypothesis generation. Future work will investigate automatic dataset weighting (to handle any statistical imbalance), integration with alternative generative models such as normalizing flows and diffusion models, and more thorough scaling studies to properly assess convergence quality and training time reduction as each dataset is further sharded. 

\section*{Acknowledgements}
This work is supported by the Scientific Discovery through Advanced
Computing (SciDAC) program via the Office of Nuclear Physics and Office of
Advanced Scientific Computing Research in the Office of Science at the U.S.
Department of Energy under contracts DE-AC02-06CH11357, 89243126CSC000213,
and DE-SC0023472, in collaboration with Argonne National Laboratory,
Jefferson Lab, National Renewable Energy Laboratory, Old Dominion
University, Ohio State University, and Virginia Tech.
This research used resources of the Argonne Leadership Computing Facility, a U.S. Department of Energy (DOE) Office of Science user facility at Argonne National Laboratory and is based on research supported by the U.S. DOE Office of Science-Advanced Scientific Computing Research Program, under Contract No. DE-AC02-06CH11357.

\clearpage

\appendix

\section{Distributed Data Parallel Training - DDP}
\label{sec:app_ddp}

In this section, we wish to briefly remind the reader of the core idea behind distributed data parallel training~\cite{torchdist2020, hvd2018}: Suppose we wish to train a neural network $M_{\phi}$ with parameters $\bm\phi$ such that it learns the functional relationship between the input features $x$ and the targets $y$. This is achieved by minimizing a loss function $\mathcal{L}$:
\begin{equation}
    \label{def_min_L}
    \phi^\ast = \mathrm{arg}\min_{\phi}\Big\{\mathcal{L}[M_{\phi}(x),y]\Big\} \implies M^\ast = M_{\phi^\ast},
\end{equation}
where $M^*$ represents the neural network with optimal parameters $\phi^{\ast}$ which fulfills: $M^*(x) = y$. The minimization in Eq.~\ref{def_min_L} is done iteratively via the parameter update:
\begin{equation}
 \label{def_update}
 \phi \mapsto \phi - \eta\cdot g_{\phi} + \xi
\end{equation}
where $g_{\phi}$ represents gradients and $\eta$ the learning rate. Depending on the optimizer (e.g. Adam, SGD,..) Eq.~\ref{def_update} might have to be modified by first and second order momenta $\xi$. The gradients $g_{\phi}$ that are needed to update the network's parameters and are determined via the average loss:
\begin{equation}
    \label{def_grad}
    g_{\phi} = \nabla_{\phi}\Big[ \frac{1}{B}\sum\limits_{j=1}^{B}\mathcal{L}(\phi,x_j,y_j) \Big]
\end{equation}
It is common practice to divide the data into batches of size $B$. A single batch then contains the inputs $x_1$,...,$x_B$ and targets $y_1$,...,$y_B$. Now assume that we divide the data within one batch even further into $k$ equal shards $\mathcal{S}_1$,...,$\mathcal{S}_k$, each of size $s$ such that: $B=ks$. One may therefore rewrite Eq.~\ref{def_grad} to:
\begin{eqnarray}
    \label{def_ddp_1}
    g_{\phi} &=& \nabla_{\phi}\Big[ \frac{1}{ks}\sum\limits_{i}^{k}\Big\{\sum\limits_{(x_j,y_j)\in\mathcal{S}_i}\mathcal{L}(\phi,x_j,y_j)\Big\} \Big] \\
    \label{def_ddp_2}
    &=& \frac{1}{k}\sum\limits_{i=1}^{k}\Big\{ \underbrace{\nabla_{\phi}\Big[\frac{1}{s}\sum\limits_{(x_j,y_j)\in\mathcal{S}_i}\mathcal{L}(\phi,x_j,y_j) \Big]}_{=g_{\phi,i}} \Big\} \\
    \label{def_ddp_3}
    &=& \frac{1}{k}\sum\limits_{i}^{k} g_{\phi,i}
\end{eqnarray}
The transition from Eq.~\ref{def_ddp_1} to~\ref{def_ddp_2} follows from the linearity of the nabla operator. Eq.~\ref{def_ddp_3} is the quintessence of distributed data parallelism, because it states that the gradient estimation, requiring one big batch $B$, can be broken up into the estimation of $k$ gradients $g_i$, each requiring a factor $k$ less data. Now the parallelism comes into play: All gradients $g_{\phi,1}$,...,$g_{\phi,k}$ can be computed in parallel as long as the aggregation in Eq.~\ref{def_ddp_3} is performed locally. 

\subsection{Requirements for DDP}
\label{sec:app_ddp_req}
The proper utilization of DDP and hence the correctness of Eq.~\ref{def_ddp_1} to~\ref{def_ddp_3} highly depends on a set of requirements that have to be fulfilled. Below is short (and incomplete) list of items a DDP practitioner needs to keep in mind~\cite{torchdist2020, hvd2018, bottou2018}:
\begin{enumerate}
    \item Each rank (or individual processing unit) that participates in DDP training has a copy of the neural network that shall be trained in parallel. Prior to training, the user needs to ensure that all ranks have the same initial network parameters and optimizer states. The former ensures that the chain from Eq.~\ref{def_ddp_1} to~\ref{def_ddp_3} is fulfilled from a computational point of view. The latter helps synchronizing the optimization step in Eq.~\ref{def_update} across all ranks. 
    \item Each rank has to have access to the accumulated gradient $g_{\phi}$ or the optimization of the local network parameters would not be correct, i.e. each rank would start seeing a different version of the network. The gradient accumulation is achieved in practice via all-reduce or ring-all-reduce.
    \item Each shard $\mathcal{S}_i$ is IID: Independent and Identically Distributed. The independency ensures that the selection of shard $\mathcal{S}_g$ is not affected by the selection of another shard $\mathcal{S}_h$. Identity ensures that the feature distributions are uniform across all shards. 
\end{enumerate}
The requirement in item 3 is not specific to DDP, as it is a general precondition for stochastic-gradient-based training to be statistically valid~\cite{bottou2018}. In the distributed setting, it applies at the level of each rank's local shard: if a shard is constructed such that certain features are under- or over-represented relative to the full population, the averaged gradient no longer constitutes a valid stochastic estimate of the true gradient, and DDP's mathematical equivalence to local training no longer holds. 

\subsection{Using DDP for distributed Multi-Dataset Analysis}
\label{sec:app_ddp_multi}
The utilization of DDP for the simultaneous analysis of multiple datasets follows naturally from the shared model framework. Let $\mathcal{L}(\phi)$ be the joint objective across all datasets:
\begin{equation}
    \label{def_L_joint}
    \mathcal{L}(\phi) = \sum\limits_{i=0}^{N-1} \mathcal{L}_i(\phi, R_i)
\end{equation}
where $\mathcal{L}_i(\phi, R_i)$ denotes the loss evaluated on dataset $R_i$. The gradients of the joint objective are then given by:
\begin{equation}
    \label{def_G_joint}
    \nabla_{\phi} \mathcal{L}(\phi) = \nabla_{\phi}\Big[ \sum\limits_{i=0}^{N-1} \mathcal{L}_i(\phi, R_i) \Big] = \sum\limits_{i=0}^{N-1}\Big\{\nabla_{\phi}\Big[\mathcal{L}_i(\phi, R_i)\Big]\Big\}
\end{equation}
The identities in Eq.~\ref{def_G_joint} hold irrespective of the joint distribution of $R_0,\ldots,R_{N-1}$, following directly from the linearity of the nabla operator. The gradient $g_{\phi,R_i}$ is locally computed on GPU $i$ and given by:
\begin{equation}
    \label{def_local_grad}
    g_{\phi,R_i} = \nabla_{\phi}\Big[\mathcal{L}_i(\phi, R_i)\Big]
\end{equation}
These local gradients are accumulated via an all-reduce mechanism:
\begin{equation}
\label{def_multi}
g_{\phi} = \frac{1}{N}\sum\limits_{i=0}^{N-1}g_{\phi,R_i}
\end{equation}
The normalization $1/N$ reflects an equal weighting of all datasets in the joint objective, ensuring that no single dataset dominates the gradient update regardless of differences in loss magnitude across datasets. One should note that this rescaling has no impact on the minimization in Eq.~\ref{def_G_joint}, since $1/N$ is a constant that does not depend on $\phi$. Please note that in practice, per-dataset weighting could be incorporated into Eq.~\ref{def_multi} by adjusting the normalization term. Unlike standard DDP, where the IID assumption ensures each $g_{\phi,R_i} $is an unbiased estimate of the same underlying gradient, here each $g_{\phi,R_i}$ provides a distinct but complementary constraint on the shared parameter space. The accumulated gradient $g_{\phi}$ therefore represents a consensus update that is consistent with all datasets simultaneously, provided all datasets are governed by the same underlying unknowns that are predicted by the model. 

\section{Teaser: Multi-Dataset Analysis in 2D}
\label{sec:app_proxy2d}

\begin{figure}[htbp]
    \centering
    \includegraphics[width=0.95\textwidth]{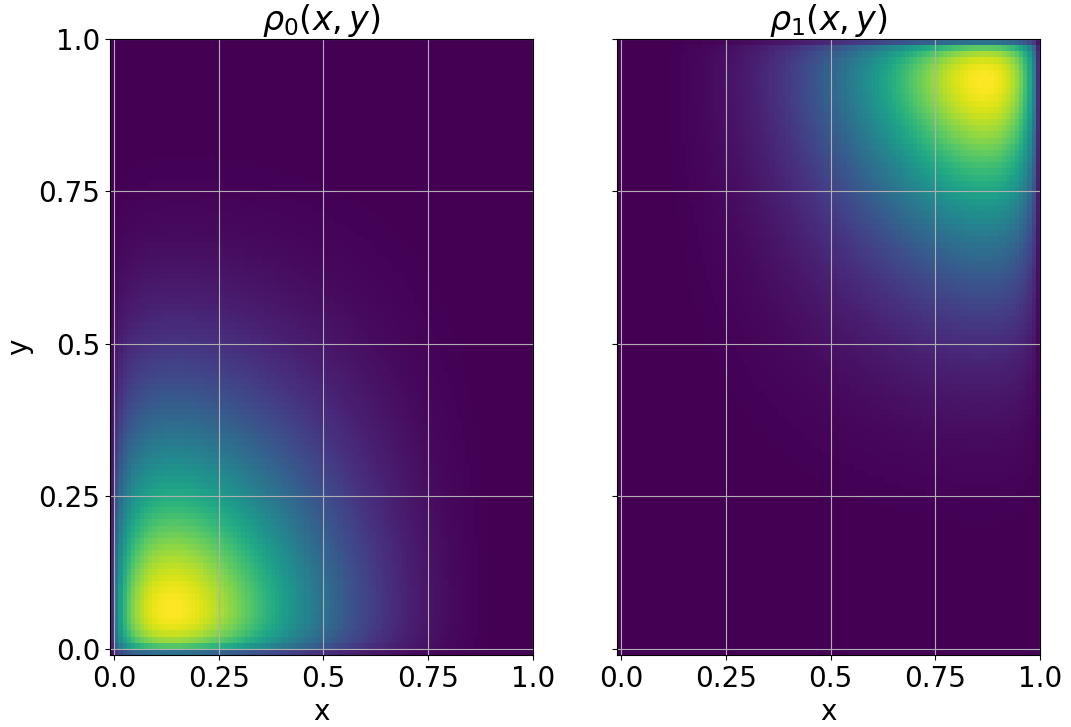} \\ 
    \vspace{0.5cm}
    \includegraphics[width=0.95\textwidth]{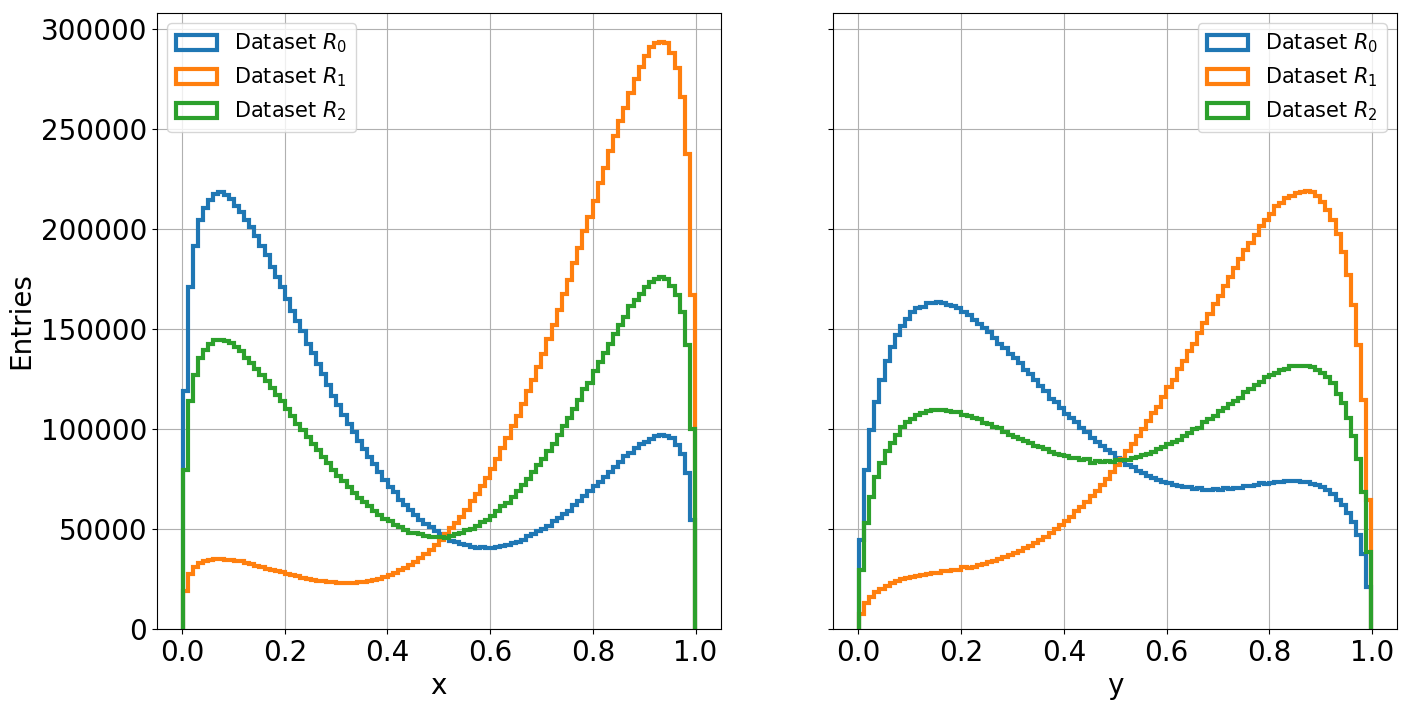}
    \caption{{\bf Top Row:} Densities $\rho_0(x,y)$ ({\bf left}) and $\rho_1(x,y)$ ({\bf right}), determined w.r.t Eq.~\ref{def_rho_0} and~\ref{def_rho_1}. Each density was created with a resolution of $100\times 100$ pixels. {\bf Bottom Row:} Datasets $R_0,R_1,R_2$ generated according to Eq.~\ref{def_proxy2d_data}.}
    \label{fig:fig_data_proxy2d}
\end{figure}

This section briefly demonstrates how the multi-dataset framework may be applied to more complex problems. The reader should note that the analysis presented here focuses on feasibility and is thus limited regarding error analysis or model tuning. We consider a problem that has been initially set up in~\cite{loits}: Two densities $\rho_0(x,y)$ and $\rho_1(x,y)$ define three datasets $R_0, R_1, R_2$:
\begin{eqnarray}
\label{def_proxy2d_data}
R_i: \{(x_k, y_k)\}_{k=1}^{N} \sim S\big(\beta_{0i}\cdot\rho_0(x,y) + \beta_{1i}\cdot\rho_1(x,y)\big)
\end{eqnarray}
where $S$ is a differentiable sampler that generates $N$ event pairs $(x_k,y_k)$ distributed according to the normalized mixture density. The coefficients $\beta_{ji}$ denote the contribution of density $\rho_j$ to dataset $i$, with: $\beta_{0i} + \beta_{1i} = 1$ for all $i$. The two densities are defined as follows:
\begin{eqnarray}
    \label{def_rho_0}
    \rho_0(x,y) &=& x^{0.5} \cdot (1-x)^{3.0} \cdot y^{0.3} \cdot (1-y)^{4.0} \cdot (1 + 0.75\cdot xy) \\
    \label{def_rho_1}
     \rho_1(x,y) &=& x^{3.0} \cdot (1-x)^{0.5} \cdot y^{4.0} \cdot (1-y)^{0.3} \cdot (1 + 0.75\cdot xy)
\end{eqnarray} 
Using the coefficients:
\begin{equation}
    \label{def_beta}
    \beta = 
   \begin{pmatrix} 
    \beta_{00} & \beta_{01} \\
    \beta_{10} & \beta_{11} \\
    \beta_{20} & \beta_{21}
   \end{pmatrix}
   =
   \begin{pmatrix} 
   0.77 & 0.23 \\ 
   0.15 & 0.85 \\ 
   0.55 & 0.45 
  \end{pmatrix}
\end{equation}
together with the two densities and following Eq.~\ref{def_proxy2d_data}, we generate three datasets with $10^7$ $(x,y)$-pairs each. A visualization of the two densities as well as the three dataset is provided in Fig.~\ref{fig:fig_data_proxy2d}.
We follow the approach introduced in section~\ref{sec:theory_dmda} and fit the datasets presented in Fig.~\ref{fig:fig_data_proxy2d} (bottom row), using a GAN together with the pipeline defined in Eq.~\ref{def_proxy2d_data}. Instead of predicting two parameters $(a_0,a_1)$, the GAN will be tasked to recover the two densities $(\rho_0(x,y),\rho_1(x,y))$.

\subsection{Setting up the GAN}
\label{sec:app_proxy2d_gan}
The datasets here consist of numerical pairs $(x,y)$, thus we keep the same discriminator architecture as introduced in section~\ref{sec:exp_setup} with the only difference that two inputs instead of one are used. The generator architecture however had to be adjusted, because the model has to predict images with two color channels (each representing one density) and a resolution matching the density images shown in the top row in Fig.~\ref{fig:fig_data_proxy2d}. The generator architecture consists of two hidden dense layers with 256 neurons each, followed by four blocks comprising a convolutional 2D layer paired with an upsampling layer. The activation function for each layer unit is chosen to be the leaky relu function with a $0.2$ slope. The final layer is another convolutional layer with a sigmoid activation function. The generator output is furthermore scaled with channel specific factors such that a physically meaningful output range is guaranteed. The number of channels for each hidden convolutional layer is $256$. The optimizers and learning rates introduced in section~\ref{sec:exp_setup} are used here as well. The generator batch size is set to one and the sampler generates $5\cdot 10^5$ events for the discriminator. Given that each datasets consists of $10^7$ events, each epoch has to process $10^7/ (5\cdot 10^5)=20$ batches. Using 500 epochs results in $10\,\rm{k}$ GAN updating steps. For this particular analysis an additional generator loss is implemented:
\begin{equation}
    \label{def_norm_loss}
    \mathcal{L}_{I} = \lambda_{I} \cdot \frac{1}{N_G} \cdot \sum\limits_{i=1}^{N_G} \Big[\Big(\frac{I_{\phi,0}-I_{\rho_0}}{I_{\rho_0}}\Big)^2 + \Big(\frac{I_{\phi,1}-I_{\rho_1}}{I_{\rho_1}}\Big)^2\Big]
\end{equation}
where $I_{\rho_k}$ represents the integral of density $\rho_k$ and $I_{\phi,k}$ denotes the integral of the corresponding predicted density. This additional loss ensures that, besides the shape, the generator model produces densities with the proper normalization. The loss scale $\lambda_{I}$ is a tunable hyper parameter and was manually tuned to be $0.008$. In a real nuclear physics analysis, the integrals $I_{\rho_k}$ would correspond to expected cross sections that one might know from past experiments, or expected theoretical values. Reproducibility and uncertainty estimation are included by applying the ensemble technique. Since this is only a feasibility study, we limit the ensemble size to three GANs in total.

\subsection{Experiments and Results}
\label{sec:app_proxy2d_exp}
\begin{figure}[htbp]
    \centering
    \includegraphics[width=1.0\textwidth]{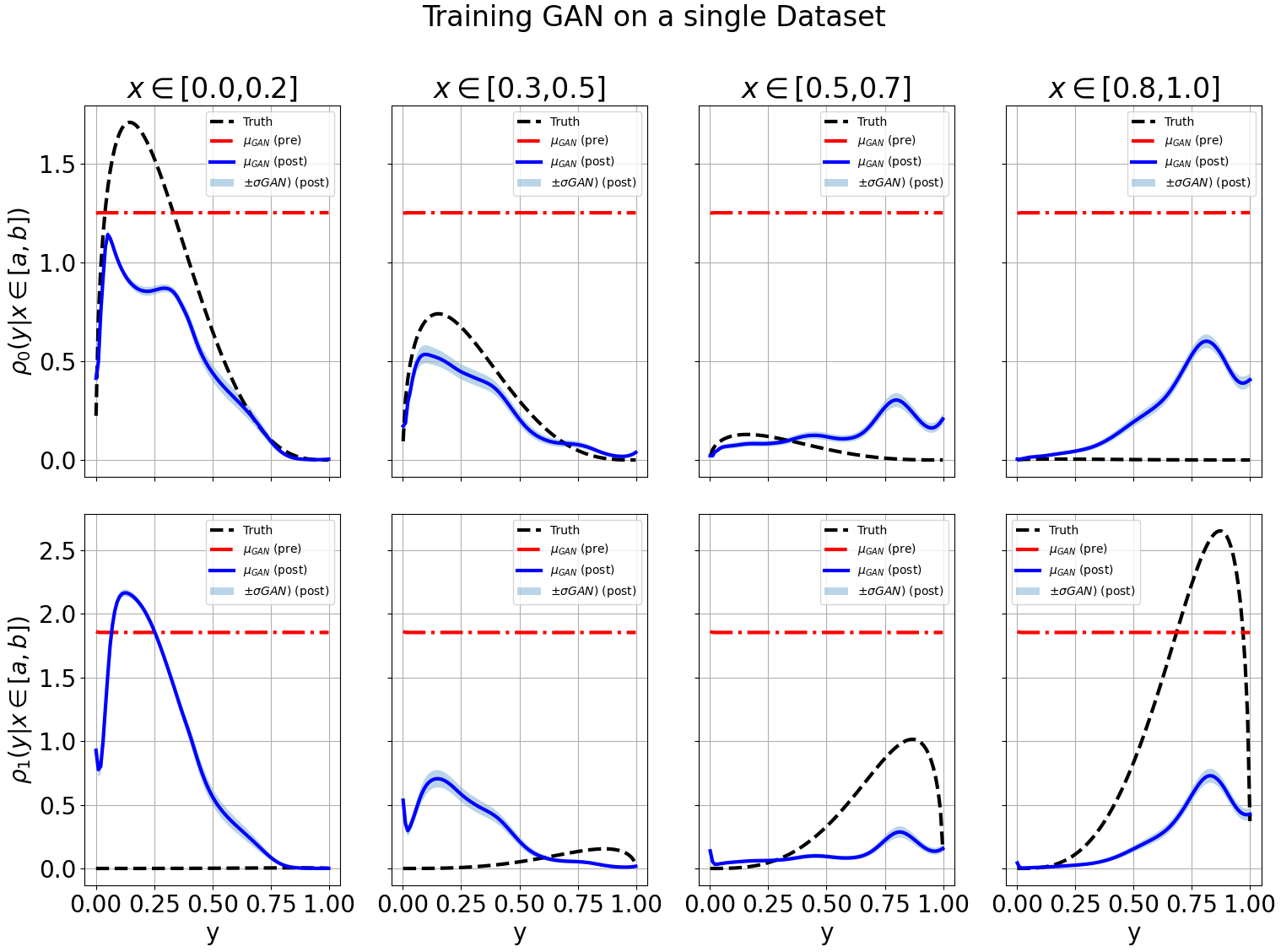}
    \caption{Density projection $\rho_0(y|x\in[a,b])$ ({\bf top row}) and $\rho_1(y|x\in[a,b])$ ({\bf bottom row}), following Eq.~\ref{def_proj_rho} for different x-intervals, as a function of y. The dashed, black lines represent the true density projections whereas the dot-dashed red and blue curves represent the mean $\mu$ of the GAN before and after training respectively. The shaded blue areas represent the uncertainty $\sigma$ of the trained GAN ensemble. The results shown here have been obtained by training the GAN on dataset $R_0$ only.}
    \label{fig:fig_proxy2d_single}
\end{figure}

\begin{figure}[htbp]
    \centering
    \includegraphics[width=1.0\textwidth]{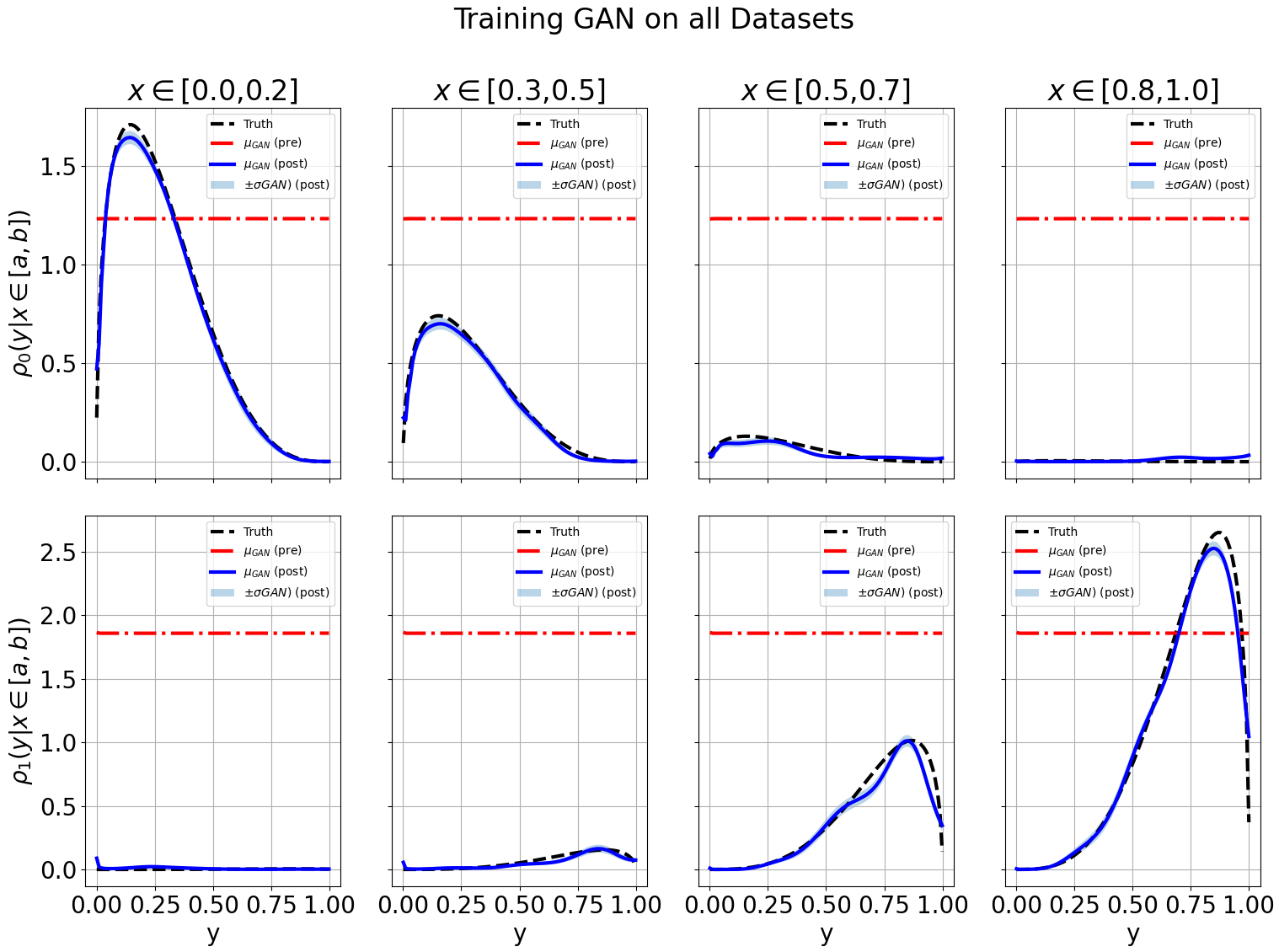}
    \caption{Density projection $\rho_0(y|x\in[a,b])$ ({top row}) and $\rho_1(y|x\in[a,b])$ ({\bf bottom row}), following Eq.~\ref{def_proj_rho} for different x-intervals, as a function of y. The dashed, black lines represent the true density projections whereas the dot-dashed red and blue curves represent the mean $\mu$ of the GAN before and after training respectively. The shaded blue areas represent the uncertainty $\sigma$ of the trained GAN ensemble. The results shown here have been obtained by training the GAN on datasets $R_0,R_1$ and $R_2$ simultaneously.}
    \label{fig:fig_proxy2d_multi}
\end{figure}

Two experiments are conducted for this teaser analysis, with the first one using dataset $R_0$ only, whereas the second experiment utilizes all three datasets $R_0,R_1,R_2$. Both experiments are run on the Jefferson lab farm (see section~\ref{sec:exp_computing}) and the results are summarized in Fig.~\ref{fig:fig_proxy2d_single} and~\ref{fig:fig_proxy2d_multi}. 
The GAN performance is assessed by monitoring projections of the densities $\rho_0$ and $\rho_1$:
\begin{eqnarray}
    \label{def_proj_rho}
    \rho_{j\in[0,1]}(y|x\in[a,b]) = \int\limits_{a}^{b}\rho_j(x,y)dx
\end{eqnarray}
The red, dot-dashed  and blue distributions in Fig.~\ref{fig:fig_proxy2d_single} compare the generator predictions before and after training. Before any parameter updates, the generator predictions are flat. Once the GAN has been exposed to data, the generator predictions show density-like features. However, a comparison to the true density projections (black, dashed curve in Fig.~\ref{fig:fig_proxy2d_single}) reveals that the model is unable to recover $\rho_0$ and $\rho_1$. With a single dataset, the system is under-determined — one mixture equation constrains two unknown densities, making their individual recovery impossible without additional information. Dataset $R_0$ is dominated by density $\rho_0$ which is why the predictions in Fig.~\ref{fig:fig_proxy2d_single} tend towards $\rho_0$ as well. 

The situation changes when the GAN is exposed to all three dataset. Fig.~\ref{fig:fig_proxy2d_multi} summarizes the corresponding results. The GAN generator learns to properly represent the input densities $\rho_0$ and $\rho_1$. While the individual projections indicate a good level of agreement. The remaining discrepancy is most pronounced in $\rho_1$ at high x, and may be reduced with increased model capacity, more training, or enhanced regularization.

\subsection{Outlook and Impact on QuantOm Project}
\label{sec:app_proxy2d_fin}
The experiments and results presented in this appendix show that the formulated multi-dataset approach is versatile and applicable to more complex pipelines. The toy problem presented here is very close to a real QuantOm application, thus demonstrating that the proposed multi-dataset framework is a suitable tool for 3D imaging of the proton. 

\clearpage

\bibliographystyle{plain} 
\bibliography{references}

\end{document}